\documentclass[natbib209]{emulateapj}

\usepackage{amsmath}
\usepackage{epsfig}

\newcommand{\pd}{\partial}
\newcommand{\oom}{\mbox{\boldmath $\Omega_0$}}
\newcommand{\bdot}{\mbox{\boldmath $\cdot$}}
\newcommand{\del}{\mbox{\boldmath $\nabla$}}
\newcommand{\curl}{\mbox{\boldmath $\nabla \times$}}
\newcommand{\dv}{\mbox{\boldmath $\nabla \bdot$}}
\newcommand{\cross}{\mbox{\boldmath $\times$}}

\newcommand{\surf}{\mbox{\boldmath ${\cal S}$}}

\newcommand{\vv}{{\bf v}}
\newcommand{\uu}{{\bf u}}
\newcommand{\BB}{{\bf B}}
\newcommand{\vort}{\mbox{\boldmath $\omega$}}

\newcommand{\DD}{\mbox{\boldmath ${\cal D}$}}
\newcommand{\grav}{\mbox{\boldmath $g$}}

\newcommand{\uvr}{\mbox{\boldmath $\hat{r}$}}
\newcommand{\uvt}{\mbox{\boldmath $\hat{\theta}$}}
\newcommand{\uvp}{\mbox{\boldmath $\hat{\phi}$}}
\newcommand{\uvl}{\mbox{\boldmath $\hat{\lambda}$}}
\newcommand{\uvz}{\mbox{\boldmath $\hat{z}$}}
\newcommand{\blam}{\mbox{\boldmath $\lambda$}}

\shorttitle{The Solar Near-Surface Shear Layer}
\shortauthors{Miesch \& Hindman}

\begin{document}

\title{Gyroscopic Pumping in the Solar Near-Surface Shear Layer}

\author{Mark S. Miesch$^{1,2}$ and Bradley W. Hindman$^2$}
\affil{$^1$High Altitude Observatory, NCAR\altaffilmark{*}, Boulder, CO, 80307-3000, USA: miesch@ucar.edu}
\affil{$^2$JILA and Department of Astrophysical and Planetary Sciences, 
University of Colorado, Boulder, CO, 80309-0440, USA}

\altaffiltext{*}{The National Center for Atmospheric Research is operated
by the University Corporation for Atmospheric Research under sponsorship
of the National Science Foundation}

\begin{abstract}
We use global and local helioseismic inversions to explore the
prevailing dynamical balances in the solar Near-Surface Shear Layer
(NSSL).  The differential rotation and meridional circulation are
intimately linked, with a common origin in the turbulent stresses of
the upper solar convection zone.  The existence and structure of the NSSL 
cannot be attributed solely to the conservation of angular momentum by solar 
surface convection, as is often supposed. Rather, the turbulent angular
momentum transport accounts for the poleward meridional flow while the
often overlooked meridional force balance is required to maintain the 
mid-latitude rotational shear.  We suggest that the base of the NSSL 
is marked by a transition from baroclinic to turbulent stresses in the 
meridional plane which suppress Coriolis-induced circulations that would
otherwise establish a cylindrical rotation profile.  The turbulent
angular momentum transport must be non-diffusive and directed radially
inward.  Inferred mean flows are consistent with the idea
that turbulent convection tends to mix angular momentum but only
if the mixing efficiency is inhomogeneous and/or anisotropic.
The latitudinal and longitudinal components of the estimated 
turbulent transport are comparable in amplitude and about
an order of magnitude larger than the vertical component.
We estimate that it requires 2--4\% of 
the solar luminosity to maintain the solar NSSL against the inertia 
of the mean flow.  Most of this energy is associated with the 
turbulent transport of angular momentum out of the layer, with 
a spin-down time scale of $\sim$ 600 days.  We also address 
implications of these results for numerical modeling of the NSSL.
\end{abstract}

\section{Introduction}\label{sec:intro}

Helioseismic inversions of global acoustic oscillation
frequencies
reveal two striking rotational boundary layers at the 
upper and lower edges of the solar convective envelope
\citep{thomp03,howe09}.  Throughout the bulk of the convection 
zone, the angular velocity $\Omega$ decreases by about
30\% between the equator and latitudes of $\pm 75^\circ$,
with conical isosurfaces such that the gradient is 
primarily latitudinal.  Substantial radial gradients
in $\Omega$ occur only near the base of the convection zone
($0.69R \lesssim r \lesssim 0.72 R$, with $R$ as 
the solar radius) where the super-adiabatic stratification
of the envelope meets the sub-adiabatic stratification of
the radiative interior, and in the surface layers
($0.95R \lesssim r \lesssim R$) where deep convection
meets the heirarchy of smaller-scale convective
motions sustained by radiative cooling in the
photosphere \citep{nordl09}.  The lower boundary
layer is known as the solar tachocline and is
thought to play an essential role in regulating the
dynamical coupling between the convection zone and the
radiative interior and in generating the large-scale
magnetic fields that underlie the solar activity cycle
\citep{hughe07b}.  The upper boundary layer, known
as the Near-Surface Shear Layer (NSSL), is similarly
complex and enigmatic.   Its existence must arise 
from nonlinear feedbacks among turbulent convective
motions spanning vastly disparate spatial and temporal 
scales with correspondingly different sensitivities
to the rotation, density stratification, and spherical
geometry.

Although somewhat less celebrated than the tachocline, the NSSL
is more accessible to helioseismic probing and as such, 
provides a unique window into the dynamics of the solar
convection zone.  The higher resolution of the helioseismic
inversion kernals near the surface allows for a more reliable
determination of the rotational gradients.  Even more significantly,
local helioseismic inversions enable a determination of the 
meridional flow throughout much of the NSSL to a degree
that is not possible in the tachocline.  Together with photospheric
Doppler and tracer measurements, such inversions indicate a
meridional flow that is highly variable but systematically
poleward \citep{snodg96,hatha96b,haber02,zhao04,gonza06,ulric10,basu10,hatha10,hatha11}.

\cite{hatha11b} has recently suggested that the dynamics that give rise
to the poleward meridional flow at the solar surface occur entirely
within the NSSL.  This conclusion is based on estimates for the subsurface
meridional flow obtained from correlation tracking of surface features.
These suggest that the poleward surface flow reverses near the base of the
NSSL, with equatorward counter-flow at a depth of 35 Mm below the photosphere
($r \approx 0.95 R$).  This is in contrast to previous estimates based
on local helioseismology, which suggest that the poleward flow 
persists throughout the upper convection zone, with the 
return flow required by mass conservatation occuring well below 0.95 $R$ 
\citep{giles97,giles99,chou01,beck02,braun98}.  If confirmed, Hathaway 
attributes this relatively shallow meridional flow structure to 
supergranulation.

More specifically, the existence of the NSSL has long been attributed to 
the tendency for photospheric convection (on scales of supergranulation 
and smaller) to conserve angular momentum locally, with fluid parcels spinning 
up and spinning down as they move toward and away from the rotation axis
respectively \citep{fouka75,gilma79b,hatha82,deros02,augus11,hatha11b}.  
Such behavior is often found in numerical simulations of rotating
convection when the rotational influence is weak, such that the
convective turnover time scale is much less than the rotation period
\citep{gilma77,gilma79b,hatha82,deros02,aurno07,augus11}.  This
condition is well satisfied in the solar surface layers where turnover
time scales associated with granulation and supergranulation are of
order a day or less and the average rotation period is 28 days.

Although the conservation of angular momentum is a valid interpretation
of the numerical experiments, it alone cannot account for the existence of 
the NSSL; there must be more to the story.  As we will demonstrate here, 
meridional forces play an essential role in determining the 
angular velocity profile in the NSSL, regardless of the nature of the 
convective angular momentum transport.  Furthermore, as we will also 
demonstrate, the angular velocity profile of the NSSL inferred from 
helioseismic inversions is not consistent with angular momentum 
homegenization.

In this paper we propose that the poleward meridional flow in the
NSSL is closely linked to the inward $\Omega$
gradient.  The physical mechanism underlying this link is
what we refer to as {\em gyroscopic pumping}, whereby a zonal forcing
(axial torque) induces a meridional flow as a consequence of
dynamical equilibration mediated by the inertia of the mean flows
(i.e.\ the Coriolis force).  In
\S\ref{sec:gp} and \S\ref{sec:meridional} we discuss the
dynamical balances that are likely to be achieved in the NSSL and 
in \S\ref{sec:helio} we exploit these dynamical balances in order to 
estimate the characteristics and efficiency of turbulent transport.  
Then, in \S\ref{sec:transport} we consider these results
from the perspective of two simple theoretical paradigms.  
We summarize our results and conclusions in \S\ref{sec:summary}.

\begin{figure}
\centerline{\epsfig{file=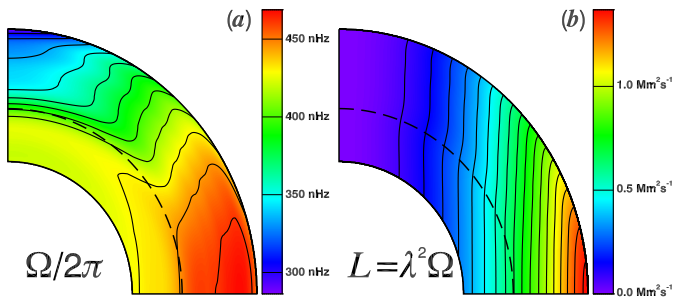,width=3.5in}}
\caption{($a$) Rotation rate $\Omega$ and ($b$) specific angular momentum 
${\cal L}$ inferred from global helioseismic inversions.  Note the sharp 
decrease in $\Omega$ near the surface which defines the NSSL.  These results
are based on RLS inversions of GONG data from four non-overlapping intervals 
in 1996, provided by R.\ Howe \citep{howe00,schou02}.
\label{fig:DR}}
\end{figure}

We wish to emphasize from the outset that we are interested in how
mean flows, namely differential rotation and meridional circulation,
respond to turbulent stresses in the solar NSSL that are currently
unknown.  Thus, in expressing the relevant dynamical balances we
will write terms involving the inertia of the mean flows (including
the Coriolis force) explicitly.  Meanwhile, we will incorporate turbulent
stresses by means of two relatively generic terms, ${\cal F}$ and ${\cal G}$, 
introduced in sections \ref{sec:concepts} and \ref{sec:zonvort} (see also 
Appendix A).  The reader who values specificity may regard ${\cal F}$ and 
${\cal G}$ as an embodiment of the convective Reynolds stress, with explicit 
expressions given in Appendix A.  This is likely to be the dominant 
contribution in the solar NSSL.  However, by leaving these terms unspecified 
we wish to highlight the generality of the physical processes we consider.
Thus, the Maxwell stress and the large-scale Lorentz force in stars and 
the artificial viscous diffusion in numerical simulations can establish
mean flows in much the same way.

\section{Gyroscopic Pumping}\label{sec:gp}

\subsection{Fundamental Concepts}\label{sec:concepts}
We begin our investigation of the dynamical balances in the NSSL
with the standard (compressible) equations of 
magnetohydrodynamics (see Appendix A), alternating throughout 
between spherical polar coordinates ($r$, $\theta$, $\phi$)  
and cylindrical coordinates ($\lambda$, $\phi$, $z$).  
Thus, $\lambda = r \sin\theta$ is the cylindrical radius
and  $z = r \cos\theta$ is the coordinate parallel
to the rotation axis (each with corresponding unit 
vectors, e.g.\ $\uvl$).
  
An equation for the conservation of 
angular momentum can be readily obtained by multiplying the 
zonal ($\phi$) component of the momentum equation by the moment
arm, $\lambda$, and then averaging over longitude 
and time (denoted by angular brackets $< >$).  This yields
\begin{equation}\label{eq:amom}
\left<\rho\right> \frac{\pd {\cal L}}{\pd t} = 
\left<\rho\right> \lambda^2 \frac{\pd \Omega}{\pd t} =
- \left<\rho \vv_m\right> \bdot \del {\cal L} + {\cal F} ~~~,
\end{equation}
where $\rho$ is the mass density and $\vv$ is the bulk velocity,
including meridional and zonal components; $\vv = \vv_m + v_\phi
\uvp$.  The term ${\cal F}$ incorporates all non-axisymmetric, 
magnetic, and viscous effects, as described further below and
in Appendix A.  Thus, we refer to it as the net axial torque.
However, if magnetic fields and viscous diffusion are neglected,
${\cal F}$ has only one component, and that is the Reynolds 
stress.  We use an 
inertial (non-rotating) coordinate system so the angular 
velocity is given by $\Omega = \left<v_\phi\right> \lambda^{-1}$ 
and the specific angular momentum is
${\cal L} = \lambda^2 \Omega$.  The meridional
velocity can be expressed in spherical and cylindrical
coordinates as 
$\vv_m = v_r \uvr + v_\theta \uvt = v_\lambda \uvl + v_z \uvz$.  

If we assume a statistically stationary state equation (\ref{eq:amom})
becomes
\cite[e.g.][]{miesc09}:
\begin{equation}\label{eq:gp}
\left<\rho \vv_m\right> \bdot \del {\cal L} = {\cal F} ~~~.
\end{equation}
Furthermore, if we assume that there is negligible mass flux 
through the solar surface, then a steady solution is only possible if
\begin{equation}\label{eq:notorque}
\int_V {\cal F} dV = 0 ~~~,
\end{equation}
where $V$ is the volume of the entire solar interior $r \leq R$.

We emphasize that equations (\ref{eq:amom}) and (\ref{eq:gp}) are 
valid for any arbitrary value of the Rossby number.  They follow
directly from the MHD equations, as demonstrated in Appendix A.
The left-hand-side includes all terms involving the inertia of
the mean flow and the right-hand-side. ${\cal F}$, is dominated 
by the Reynolds stress.

More specifically, the left-hand-side of equation (\ref{eq:gp}) 
repesents the advection of
angular momentum by the mean meridional circulation and the
right-hand-side incorporates all other forces, most notably the
convective Reynolds stress and the Lorentz force.  Molecular viscosity
can also contribute to the net torque ${\cal F}$ but this is expected
to be negligible in stars. Any imbalance between the forces that
contribute to ${\cal F}$ will induce a meridional flow $\left<\rho
\vv_m\right>$ across ${\cal L}$ isosurfaces.  Although $\Omega$
contours are nearly radial in the bulk of the solar convection zone,
${\cal L}$ contours are nearly cylindrical, as demonstrated in Figure
1, increasing away from the rotation axis such that
\begin{equation}\label{eq:cylindrical}
\del {\cal L} \approx \frac{\pd {\cal L}}{\pd \lambda} ~ \uvl 
\end{equation}
and $\pd {\cal L}/\pd \lambda > 0$.  Thus, equation (\ref{eq:gp}) 
implies that a retrograde torque ${\cal F} < 0$ will induce 
a flow toward the rotation axis while a prograde torque 
${\cal F} > 0$ will induce a flow away from the rotation axis.  
This is the concept of gyroscopic pumping discussed by 
McIntyre (1998, 2007; see also Haynes et al.\ 1991, 
Garaud \& Acevedo Arreguin 2009, and Garaud \& Bodenheimer 2010).
\nocite{hayne91,mcint98,mcint07,garau09,garau10}
Thus, we refer to equation (\ref{eq:gp}) as the gyroscopic pumping
equation.

The process by which equation (\ref{eq:gp}) is established will be
addressed in \S\ref{sec:meridional}. Here we focus on its
implications.   To proceed, we assume that the mass flux is
divergenceless, $\dv \left<\rho \vv_m\right> = 0$.  This can
be justified with the anelastic approximation but that is not
necessary; a non-divergent mean mass flux follows directly
from the compressible mass continuity equation under the
assumption of a statistically stationary state.  We may then
define a streamfunction $\Psi$ such that
\begin{equation}\label{eq:psidef}
\left<\rho v_\lambda\right> = \frac{\pd \Psi}{\pd z}
\mbox{\hspace{.1in} and \hspace{.1in}}
\left<\rho v_z\right> = - \frac{1}{\lambda} \frac{\pd }{\pd \lambda} \left(\lambda \Psi\right) ~~~.
\end{equation}

If the ${\cal L}$ contours are cylindrical, as expressed in equation (\ref{eq:cylindrical}),
then $\Psi$ follows directly from equations (\ref{eq:gp}) and (\ref{eq:psidef}):
\begin{equation}\label{eq:psisoln}
\Psi(\lambda,z) = \left(\frac{d {\cal L}}{d\lambda}\right)^{-1} 
\int_{z_b}^z {\cal F}(\lambda,z^\prime) dz^\prime
\end{equation}
where $z_b = (R^2 - \lambda^2)^{1/2}$.  In obtaining equation (\ref{eq:psisoln})
we have assumed that there is no mass flux through the photosphere
so $\Psi = 0$ at $r = R$.  Thus, $\Psi$ is obtained by integrating along
a cylindrical surface, beginning at the photosphere in the northern 
hemisphere ($z = z_b$) and proceeding in the negative $z$ direction, 
through the equatorial plane, and continuing to the photospheric boundary 
in the southern hemisphere ($z = - z_b$).

An appreciation for how gyroscopic pumping operates is best obtained
by considering the simplest case in which the differential rotation is
weak.  This limit is satisfied if $R_o^{DR} << 1$, where 
$R_o^{DR} = \Delta \Omega/(2 \Omega)$ is the Rossby number 
associated with the differential rotation, and $\Delta \Omega$ is 
some measure of the variation of the rotation rate with latitude 
and radius (a rigorous derivation of the limit yields
$\Delta \Omega = \lambda \left\vert \del \Omega \right\vert$). 
This criterion is approximately satistied in the Sun; 
helioseismic inversions indicate $R_o^{DR} \sim 0.16$.  In
the low Rossby number limit, the uniform rotation
component $\Omega_0$ dominates the angular momentum gradient and
$d{\cal L}/d\lambda = 2 \lambda \Omega_0$.  Substituting this into 
(\ref{eq:psisoln}) then yields $\Psi$ for a given ${\cal F}$.

This simple example emphasizes an important point: Here, the 
net {\em zonal} force (encompassed in the net axial torque ${\cal F}$) 
determines the {\em meridional} flow, {\em not} necessarily the 
differential rotation.  
For a given ${\cal F}$, equation (\ref{eq:gp}) provides 
a direct link between the meridional flow and the turbulent
angular momentum transport, valid to lowest order in the 
Rossby number (which is assumed to be small).  The differential
rotation is determined by other mechanisms, as discussed in
\S\ref{sec:clarify}, \S\ref{sec:meridional}, and \S\ref{sec:transport}.
An idealized demonstration of how this works is given in
Appendix B.

Note also that a cylindrical torque ${\cal F} = {\cal F}(\lambda)$
is ruled out by equation (\ref{eq:psisoln}).  This would imply
a cylindrical mass flux $\left<\rho v_\lambda\right>$ that is
independent of $z$, which is in turn ruled out by mass conservation
and our requirement that there be no flow through the surface
$r = R$.  More explicitly, we can say that if the ${\cal L}$ 
profile is cylindrical [eq.\ (\ref{eq:cylindrical})], then 
a steady meridional flow is only possible if 
$\int_{z_b}^{-z_b} {\cal F} dz = 0$.  In other words, 
the meridional flow responds mainly to the
axial variation of the torque, $d{\cal F}/dz$.  The amplitude 
of the resulting $\vv_m$ is proportional to ${\cal F}$ and inversely 
proportional to $\Omega_0$; For a given rotation rate, the stronger 
the zonal force, the stronger the meridional flow that is induced.

For finite values of $R_o^{DR}$ equation (\ref{eq:gp}) must still hold
in a steady state but the contribution of the differential rotation
to $\del {\cal L}$ cannot be neglected.  The meridional circulation
will redistribute angular momentum so ${\cal L}$ will depend on
$\Psi$ and the gyroscopic pumping equation is nonlinear even if
${\cal F}$ is fixed.  Linear and nonlinear feedbacks of 
$\Omega$ and $\Psi$ on ${\cal F}$ complicate the problem further.  
A unique solution requires consideration of the meridional momentum 
and energy equations, as well as mean flow profiles that adjust in 
order to minimimize the net axial torque ${\cal F}$.  We address 
these issues in \S\ref{sec:clarify} and \S\ref{sec:meridional} below.

\subsection{Clarifications and Generalizations}\label{sec:clarify}

As emphasized in the last paragraph of \S\ref{sec:concepts}, the
gyroscopic pumping equation (\ref{eq:gp}) is robust.  It holds
for any arbitrary value of the Rossby number provided there
exist well-defined, persistent mean flows.  Furthermore, 
$\del {\cal L}$ is undeniably directed away from the rotation
axis in the NSSL, as revealed by global helioseismic
inversions (Fig.\ \ref{fig:DR}$b$).  Thus, the sense and amplitude
of the meridional flow is linked to the net axial torque 
${\cal F}$.  However, as also noted in \S\ref{sec:concepts},
this link is in general nonlinear and depends on factors other
than the angular momentum transport.  In this section we discuss
some of the subtlties involved.

We begin with equation (\ref{eq:psisoln}), which rests on the 
approximation of a cylindrical angular momentum profile,
as expressed in equation (\ref{eq:cylindrical}).  Although
this is an instructive example with relevance to the Sun
(cf.\ Fig.\ \ref{fig:DR}$b$), the concept of gyroscopic pumping
is more general and applies straightforwardly to other scenarios 
as well.  In general, the meridional flow through ${\cal L}$ isosurfaces 
is linked to the net axial torque ${\cal F}$ while the
flow along ${\cal L}$ isosurfaces follows from mass conservation.
The integral in equation (\ref{eq:psisoln}) for $\Psi$ would 
then proceed along ${\cal L}$ isosurfaces.

Another issue mentioned in \S\ref{sec:concepts} concerns the
dependence of the net axial torque ${\cal F}$ on the mean
flows and the possibility that the mean flows adjust to 
minimize ${\cal F}$.  In some circumstances, this dependence
may be regarded in terms of a differential operator that 
operates on the mean rotation profile: 
${\cal F} = {\cal F}\left\{\Omega\right\}$.  An 
example is the case of turbulent diffusion, in which
the angular momentum flux is proportional to $\del \Omega$,
as discussed in \S\ref{sec:transport}.  In this case,
the gyroscopic pumping equation (\ref{eq:gp}), admits a
homogeneous solution as well as a particular solution.
In other words, we may write $\Omega = \Omega_h + \Omega_p$
where
\begin{equation}\label{eq:homo}
{\cal F}\left\{\Omega_h\right\} = 0
\mbox{\hspace{0.1in} and \hspace{0.1in}}
{\cal F}\left\{\Omega_p\right\} = - \left<\rho \vv_m\right> \bdot \del {\cal L}  ~~~.
\end{equation}
This decomposition is only rigorously valid if the operator is linear
and if the dependence of the angular momentum advection $\left<\rho \vv_m\right> {\cal L}$
on $\Omega$ is linearized in some way (e.g.\ for $R_o^{DR} << 1$).  Even so, 
it serves to illustrate intuitively how the nature of ${\cal F}$ has 
implications for the differential rotation as well as for the meridional 
circulation.  Even in the absence of meridional flow, the solution to the 
homogeneous equation ${\cal F} = 0$ may in general exhibit a differential 
rotation that depends on the nature of the operator and the boundary conditions.  
Examples are given in \S\ref{sec:homo}.   

Yet, the basic premise of gyroscopic pumping is still valid, namely
that a net axial torque ${\cal F} \neq 0$ can only be sustained in
a steady state if angular momentum is continually replenished by 
advection from the surrounding fluid.  The meridional flow, differential 
rotation, and turbulent stresses will adjust until this is achieved.  
Meridional and zonal forces both play a role in this nonlinear dynamical 
adjustment as described in \S\ref{sec:thex} and contribute to the mean 
flow profiles that are ultimately realized.   Still, the nearly 
cylindrical ${\cal L}$ profile revealed by helioseismology 
(Fig.\ \ref{fig:DR}$b$) provides
a robust link between the meridional flow and the net zonal forcing, 
expressed in equation (\ref{eq:gp}).  Furthermore, the prominent
axial $\Omega$ gradient $\pd \Omega/\pd z$ revealed by helioseismic
inversions provides a robust link between the rotational shear and
the meridional forcing, as addressed in \S\ref{sec:meridional}.
Such non-intuitive links between meridional/zonal flows and zonal/meridional 
forcing are mediated by the Coriolis force so they are most prominent 
in rapidly rotating systems.  However, their robustness applies 
even in the solar NSSL where the Rossby number based on convection 
is large (\S\ref{sec:implications}).

Another important point about gyroscopic pumping is that it is
inherently a {\em non-local} process. A localized torque ${\cal F}$ will
in general induce a global circulation, extending far beyond the
forcing region \citep{hayne91}.  This has particular significance with
regard to the problem of tachocline confinement, whereby gyroscopic pumping in the
convection zone induces a circulation that burrows downward into the
radiative interior with time unless other physical processes suppress
it \citep{spieg92,gough98,garau08,garau09,garau10}.  Likewise, zonal 
forces in the NSSL have potential implications for mean meridional 
and zonal flows throughout the convection zone.  For a demonstration
of how a local zonal force in the NSSL can induce a global meridional
flow, see the analytic example presented in Appendix B.

Yet, the coupling between the bulk of the convection zone (CZ) and the
NSSL will also work the other way.  Namely, there must be a net torque
in the convection zone ${\cal F}_{CZ}$ that induces a meridional flow
by gyroscopic pumping that will extend into the NSSL.  Furthermore,
given the relatively large mass and energy content of the CZ relative
to the NSSL, we may expect this meridional flow to overwhelm that
which is driven by net axial torques within the NSSL, ${\cal F}_{NSSL}$.
In order to account for the solar differential rotation, the sense of
the deep-seated angular momentum transport must be such that ${\cal
F}_{CZ}$ is positive at low latitudes and negative at high latitudes.
This will induce a counter-clockwise circulation in the northern
hemisphere via eq.\ (\ref{eq:gp}) that will pump mass flux into the
NSSL, maintaining a poleward flow even in the absence of any
turbulent stresses within the NSSL itself.  However, if ${\cal
F}_{NSSL}$ were indeed zero, then this CZ circulation would
redistribute angular momentum until the ${\cal L}$ contours aligned
with the streamlines of the meridional flow such that $\left<\rho
\vv_m\right> \bdot \del {\cal L} = 0$.  This is clearly not the case
in the Sun, where the meridional flow is poleward and the ${\cal L}$
contours are nearly cylindrical (Fig.\ \ref{fig:DR}$b$).  The
meridional flow clearly crosses ${\cal L}$ isosurfaces, 
so ${\cal F}_{NSSL}$ must be nonzero.

In short, gyroscopic pumping by net axial torques in the deep convection
zone may contribute to the poleward flow in the NSSL but angular
momentum transport within the NSSL itself must also play a role.
This is consistent with the mean-field simulations of \cite{rempe05}
who considered the maintenance of mean flows in the solar interior
based on idealized parameterizations for the turbulent momentum and
energy transport.  He found poleward meridional flow near the 
solar surface even without a NSSL.  When he included an inward
(cf.\ \S\ref{sec:anisotropy}) angular momentum transport in 
a thin layer near the surface, he found both an NSSL-like 
shear layer where $\pd \Omega /\pd r < 0$ as well as an 
enhancement of the poleward flow.

\subsection{Implications}\label{sec:implications}

In light of the discussion in \S\ref{sec:concepts} and
\S\ref{sec:clarify}, the implications of equation (\ref{eq:gp}) for
the solar NSSL are clear.  The same physical mechanism responsible for
the deceleration of the rotation rate in the solar surface layers
inferred from helioseismology ($\pd \Omega/\pd r < 0$) is also 
responsible for the poleward meridional flow inferred from helioseismic 
and Doppler measurements ($\left<v_\theta\right> < 0$ in the northern
hemisphere).  The underlying cause of both phenomena is a retrograde net
axial torque
(zonal force) ${\cal F}$ which is most likely due to a divergence in
the angular momentum transport by the convective Reynolds stress.  Its
localization near the solar surface (resulting in a large $\pd {\cal
F}/{\pd z}$ that effectively generates a meridional flow) is in turn a
likely consequence of the rapidly changing length and time scales of
convection, $L_c$ and $\tau_c$.  The influence of rotation on convection
is typically quantified by the Rossby number based on the convective
time scales $R_o = (2 \Omega \tau_c)^{-1}$ (as opposed to $R_o^{DR}$ above).  
Estimates for giant cells ($\tau \sim $ 10--20 days, $R_o
\sim$ 0.1--0.2) and granulation ($\tau \sim 8$ min, $R_o \sim 400$) suggest
that $R_o$ should cross unity somewhere in the vicinity of the lower
NSSL, likely signifying a qualitative change in convective transport.

Thus, in this section we have established that a net retrograde zonal
force ${\cal F} < 0$ in the NSSL will induce a poleward meridional
flow.  This is a very general result, independent of the nature of
${\cal F}$, which we address in \S\ref{sec:transport}.  Before
proceeding to this, however, we discuss another general and important
maxim; a net axial torque ${\cal F}$ cannot in itself account for the
existence of the NSSL.  Again, this conclusion is completely
independent of the nature of ${\cal F}$ and follows essentially from
the Taylor-Proudman theorem.  Coriolis-induced meridional flows will
tend to eliminate all axial rotational shear $\pd \Omega / \pd z$ 
{\em regardless} of the amplitude and structure of the zonal
forcing.  Thus, the axial shear $\pd \Omega / \pd z$ in the NSSL 
must be determined not by convective angular momentum transport 
${\cal F}$ but rather by turbulent and baroclinic stresses in 
the meridional plane, which we now address (\S\ref{sec:meridional}).

\section{Meridional Force Balance}\label{sec:meridional}

\subsection{The Zonal Vorticity Equation}\label{sec:zonvort}

In \S\ref{sec:concepts} we discussed how a net axial torque
${\cal F}$ can induce a meridional circulation through the gyroscopic
pumping equation (\ref{eq:gp}) but we said little about how this 
balance is achieved.  We also left open the question of how the
differential rotation profile is established, particularly
in the low Rossby number limit ($R_o^{DR} << 1$) when ${\cal F}$
is directly linked to the meridional flow through 
eq.\ (\ref{eq:psisoln}).

To address these issues for any arbitrary value of the Rossby number,
we must also consider the meridional components of the momentum equation.  Exploiting
the divergenceless nature of the mean mass flux $\left<\rho \vv_m\right>$, 
we can combine these two equations into one by considering the zonal 
component of the curl; in other words, the zonal vorticity equation, averaged
over longitude and time.  In order to illustrate how the system adjusts in
response to specified forcing scenarios, we will temporarily retain the 
time derivative and write this equation as follows:
\begin{equation}\label{eq:mer}
\frac{\pd}{\pd t} \left<\omega_\phi\right> = 
\lambda \frac{\pd \Omega^2}{\pd z} + {\cal B} + {\cal G} ~~~,
\end{equation}
The vorticity is defined as $\vort = \curl \vv$, with $\omega_\phi$ as
the zonal component.   We emphasize that equation (\ref{eq:mer}) is valid 
for any arbitrary value of the Rossby number, as demonstrated in Appendix A.
The first term on the right-hand side includes the complete inertia of the
mean zonal flow, including uniform (Coriolis) and differential 
rotation components.  This is followed by the baroclinic term
\begin{equation}\label{eq:bc}
{\cal B} \equiv \frac{\del \left<\rho\right> \cross \del \left<P\right>}{\left<\rho\right>^2}
\bdot \uvp ~~~,
\end{equation}
where $\rho$ is the density as before and $P$ is the pressure.  
The final term on the right-hand-side of equation (\ref{eq:mer}), ${\cal G}$,
represents turbulent stresses in the meridional plane.  As with the 
net axial torque ${\cal F}$, ${\cal G}$ includes the Reynolds stress, the
Lorentz force, and the viscous diffusion.  We have also included in
${\cal G}$ baroclinic contributions associated with thermal
fluctuations, e.g.\ $\rho^\prime = \rho - \left<\rho\right>$.  
An explicit expression for ${\cal G}$ is given in Appendix A,
along with the derivation of equation (\ref{eq:mer}).

\subsection{A Thought Experiment}\label{sec:thex}

We now describe a simple thought experiment in order to illustrate
several important points about gyroscopic pumping and to gain an
intuitive feel for how it operates within the context of the NSSL.
We stress that this is an idealized example intended to illustrate
fundamental physical principles, namely that some meridional 
forcing is needed to account for the existence of the NSSL.
We then proceed to discuss the nature of this meridional forcing
in \S3.3.

Consider a rotating spherical volume of radius $R$ subject to 
a specified net axial torque ${\cal F}$ that turns on at some 
instant $t = t_0$.  In analogy to the NSSL, we will
assume that ${\cal F} < 0$ in a thin layer near the surface,
say $r_s \leq r \leq R$.  To be explicit, we can set $r_s \approx 0.95 R$
(see \S\ref{sec:homo}).
In order to allow the system to eventually reach an equilibrium 
state, we assume that the net (integrated) torque in the convection 
zone is positive, such that ${\cal F}$ satisfies
equation (\ref{eq:notorque}).

For simplicity we will assume that the baroclinic and 
turbulent stresses in the meridional plane vanish, 
so ${\cal B} = {\cal G} = 0$.  This corresponds to a flow
that is axisymmetric, non-magnetic, non-diffusive, and
adiabatic.  Furthermore, we assume that ${\cal F} = 0$ 
for $t < t_0$.  Thus, before time $t_0$, the system can 
sustain an initial equilibrium state with a cylindrical rotation 
profile $\Omega = \Omega_i(\lambda)$ and no meridional flow, 
$\vv_m  = 0$ (satisfying the Taylor-Proudman theorem; see, 
e.g.\ Pedlosky 1987).
In analogy with the Sun, we assume that the equator rotates 
faster than the poles, so $d\Omega_i/d\lambda > 0$.  It 
follows also that the angular momentum increases outward, 
$\del {\cal L} \bdot \uvl > 0$ and is cylindrical
($\del {\cal L} \bdot \uvz = 0$), as expressed in 
equation (\ref{eq:cylindrical}).  This initial 
equilibrium satisfies equations (\ref{eq:gp}) and
(\ref{eq:mer}), with $\pd \left<\omega_\phi\right>/\pd t = 0$.

Now turn on the torque ${\cal F}$ at $t = t_0$.  How
does the system respond?  Initially, the torque will
slow down the rotation rate in the NSSL, tilting the
$\Omega$ contours away from the rotation axis.
This will produce an axial rotation gradient $\pd \Omega /\pd z$
which will in turn induce a counter-clockwise meridional
circulation in the northern hemisphere (negative 
$\left<\omega_\phi\right>$) according to 
equation (\ref{eq:mer}).  In other words, there will be
a poleward flow in the NSSL and a net equatorward return
flow in the deeper convection zone to ensure mass conservation.

This induced meridional flow will redistribute angular momentum,
altering the $\Omega$ profile.  In the NSSL, angular momentum
will be advected from low latitudes toward the poles, accelerating
the rotation rate, accompanied by a deceleration of the deeper
convection zone.  This will proceed until a new, final 
equilibrium is reached, again with a cylindrical rotation profile
$\Omega = \Omega_f(\lambda)$.  The new profile will be different 
from the initial profile ($\Omega_f \neq \Omega_i$), with
cylindrical isorotation surfaces shifted away from the
rotation axis (positive $\uvl$ direction).  Furthermore, 
the torque ${\cal F}$ will sustain a steady, nonzero meridional 
circulation $\Psi_f$, given by equation (\ref{eq:psisoln}),
with poleward flow in the NSSL and an equatorward return flow 
in the deep CZ.

In summary, the response of the system to a retrograde zonal torque
${\cal F}$ localized in the NSSL is to establish a poleward flow that
will increase in amplitude until the prograde angular momentum advected 
into the layer from the deep CZ balances the retrograde angular momentum 
imparted by the torque.  The meridional flow is predominantly poleward 
($\vv_m \approx v_\theta \uvt$) rather than cylindrically inward 
($v_\lambda \uvl$) because of the thin radial extent of the NSSL 
(see Appendix B for a demonstration).  Meanwhile, the rotation
profile is altered but still cylindrical, $\Omega = \Omega_f(\lambda)$.

This simple thought experiment illustrates two important points.
First, {\em gyroscopic pumping is mediated by the Coriolis force but
the two phenomena are not equivalent}.  The meridional components 
of the Coriolis force can in principle vanish (for the special case
of $\Omega_f$ = constant) while the system still sustains a gyroscopically-pumped
meridional circulation.  More generally, for any cylindrical rotation
profile $\Omega(\lambda)$, the meridional components of the Coriolis
force are opposed by pressure gradients (geostrophic balance) so 
they impart no zonal vorticity and thus no meridional momentum.
The meridional flow is determined not by the Coriolis force (which
responds to $\Omega$), but rather by the conservation of angular 
momentum (which responds to ${\cal F}$).

The second point is even more important: {\em The mere existence 
of a retrograde torque ${\cal F}$ does not guarantee the 
presence of a near-surface shear layer}.  As stated in the 
introduction, even if the concept is valid, angular momentum
conservation by turbulent convection alone cannot account for 
the existence of the NSSL.  In the absence of baroclinic, 
turbulent, or magnetic stresses in the meridional plane 
(${\cal B} = {\cal G} = 0$), the meridional 
circulation will wipe out all axial shear $\pd \Omega / \pd z$,
{\em regardless of the nature, magnitude or profile of the 
convective angular momentum transport} (embodied by ${\cal F}$).  

The robustness of cylindrical rotation profiles is a consequence
of the Taylor-Proudman theorem \citep[e.g.][]{pedlo87}.  No matter
how strong ${\cal F}$ is, a commensurate meridional circulation 
will suppress axial shear on a time scale comparable to the 
rotation period ($\sim$ 28 days).  The prominent axial shear 
$\pd \Omega / \pd z$ exhibited by helioseismic rotation 
inversions (Fig.\ \ref{fig:DR}$a$) then implies that some 
meridional forcing, either baroclinic or ``turbulent'', 
${\cal B}$ or ${\cal G}$, is necessary to account for 
the existence of the NSSL as well as the detailed structure
of the inferred $\Omega$ profile.

\subsection{What Determines the Rotational Shear in the NSSL?}\label{sec:what}

In this section we argue that the existence and location of the NSSL 
can be attributed to a qualitative change in the meridional force balance,
marked by a transition from baroclinic to turbulent stresses; ${\cal B}$
to ${\cal G}$.  Furthermore, it is the meridional stresses (${\cal G}$ 
and/or ${\cal B}$) rather than the angular momentum transport that 
largly determine the slope of the $\Omega$ profile (although the 
negative sign of $\pd \Omega / \pd r$ still requires a retrograde
net axial torque ${\cal F} < 0$).

Although there are still some subtle uncertainties regarding the
underying dynamics, recent global solar convection simulations and
mean-field models have converged on a consistent paradigm whereby the
nearly radial angular velocity contours (conical $\Omega$ isosurfaces)
at mid-latitudes in the deep CZ are attributed to baroclinic forcing
\begin{equation}\label{eq:tw}
\frac{\pd \Omega^2}{\pd z} = - \frac{{\cal B}}{\lambda}
\approx \frac{g}{r \lambda C_P} \frac{\pd \left<S\right>}{\pd \theta}
\mbox{\hspace{.5in} (deep CZ)}
\end{equation}
where $g$ is the gravitational acceleration, $S$ is the specific entropy, 
and $C_P$ is the specific heat at constant pressure
\citep{kitch95,ellio00,robin01,brun02,rempe05,miesc06,brun11}.  This
is referred to as thermal wind balance \citep{pedlo87,miesc09}.  The
convective Reynolds stress is still necessary to account for the
amplitude and sense of the angular velocity contrast between equator
and pole, $\Delta \Omega$, but without baroclinicity, models generally
yield cylindrical (Taylor-Proudman) profiles, in contrast to the conical 
profiles inferred from helioseismic inversions (Fig.\ \ref{fig:DR}$a$).  
In early models, the latitudinal entropy gradient in equation (\ref{eq:tw}) 
is established entirely by a latitude-dependent convective heat flux
attributed to the influence of rotation on convective motions
\citep{kitch95,ellio00,robin01,brun02}.  However, recent models have
demonstrated that thermal coupling to the subadiabatic portion of the
tachocline by means of a gyroscopically-pumped meridional circulation
may contribute to establishing the requisite thermal gradients 
\citep{rempe05,miesc06,brun11}.

In any case, the nature and prominence of the baroclinic term in
convection simulations and mean-field models hinges on the low
effective Rossby number in the deep CZ.  By {\em effective} Rossby
number we are referring to the amplitude of the meridional components
of the convective Reynolds stress relative to the deflection of the
differential rotation by the Coriolis force and the baroclinic
forcing.  However, as noted in \S\ref{sec:implications}, the Rossby number
rises steadily with radius in the convection zone, becoming much
greater than unity in the NSSL.  This alone suggests that the
turbulent stresses, represented by ${\cal G}$, may overwhelm the
baroclinic term ${\cal B}$ in the solar surface layers.  {\em In fact,
this shift in the meridional force balance (from ${\cal B}$ to
${\cal G}$) may mark the base of the NSSL even more significantly
than a change in the turbulent angular momentum transport ${\cal F}$.}

Thus, we propose that in the NSSL thermal wind balance,
equation (\ref{eq:tw}) is replaced by
\begin{equation}\label{eq:CG}
\frac{\pd \Omega^2}{\pd z} = - \frac{{\cal G}}{\lambda}
\mbox{\hspace{.5in} (NSSL)}.
\end{equation}
The implication would then be that {\em turbulent stresses in the meridional 
plane, ${\cal G}$, largely determine the differential rotation profile 
in the NSSL}.  This hypothesis is consistent with global simulations
of solar convection which, although they do not yet reproduce the
NSSL, do exhibit a departure from thermal wind balance in the
surface layers above $r \sim 0.95$ \citep{ellio00,brun10,brun11}.
This departure is attributed to a combination of resolved convective
Reynolds stresses and modeled sub-grid scale diffusion, both 
captured in our turbulent transport term ${\cal G}$.  Achieving
a NSSL in global convection simulations will likely require
a reliable representation of turbulent transport by 
unresolved (subgrid-scale) motions, as discussed in 
\S\ref{sec:gp} and \S\ref{sec:transport}.

Can solar observations and models reveal what
the meridional force balance is in the NSSL, or in other 
words, which is bigger, ${\cal B}$ or ${\cal G}$?  In principle 
the answer is yes, but as we discuss in the remainder of
this section, current results are inconclusive.

Helioseismic rotational inversions provide a good estimate for
the left-hand-side of equation (\ref{eq:tw}), namely 
${\cal C} = -\lambda \pd \Omega^2/\pd z$ (\S\ref{sec:helio}).
If we had a reliable estimate for the right-hand-side, ${\cal B}$,
from solar observations and if its amplitude were $<< {\cal C}$, 
then we could rule out equation (\ref{eq:tw}) and thereby verify 
the proposed balance expressed in equation (\ref{eq:CG}).  

One way to estimate the amplitude of ${\cal B}$, at least in
principle, is by means of helioseismic structure inversions.
The thermal gradients that give rise to ${\cal B}$ imply 
aspherical sound speed variations.  If these gradients are to 
balance the axial shear in the rotation rate through baroclinic 
torques as expressed in equation (\ref{eq:tw}), then their relative 
amplitude must be of order
\begin{equation}\label{eq:Svar}
\frac{\Delta S}{C_p} \sim
\frac{2}{\pi C_P} \frac{\pd \left<S\right>}{\pd \theta} \sim 
\frac{2 r \lambda}{\pi g} \frac{\pd \Omega^2}{\pd z} \sim 2\times 10^{-5}
\end{equation}
where $\Delta S$ is the entropy variation from equator to pole.
The numerical estimate in equation (\ref{eq:tvar}) is based on 
global rotational inversions as described in \S\ref{sec:helio} 
below.

Such subtle thermal variations are beyond the detection limit
of current helioseismic structure inversions, even in the NSSL 
\citep{gough96b,christ02}.  \cite{brun10} have reported 
aspherical sound speed variations much larger than this,
of order $10^{-5}$--$10^{-4}$, but the amplitude of the variations 
peaks near the base of the convection zone at high latitudes, where
the reliability of the inversions is questionable.  If these inversions
are indeed valid and if they do indeed trace thermal gradients as 
opposed to magnetic effects, then thermal wind balance, equation
(\ref{eq:tw}) would not be satisfied and turbulent stresses ${\cal G}$ 
would have to contribute to the meridional force balance throughout
the deep convection zone as well as the NSSL.  More work is needed
to set stricter limits on aspherical sound speed variations from 
helioseismic structure inversions and whether they are consistent
with helioseismic rotational inversions within the context of
thermal wind balance.  For now we regard the issue as not 
yet settled.

An independent estimate for ${\cal B}$ near the solar surface can
(again, in principle) be obtained from photospheric irradiance 
measurements.  The final equality in equation (\ref{eq:tw}), expressing
${\cal B}$ in terms of the latitudinal entropy gradient, applies for an 
ideal gas equation of state and a nearly hydrostatic, adiabatic stratification
\citep[e.g.][]{miesc09}.  An alternative expression can be obtained by
assuming that the photosphere is an isobaric surface.  In this case the
baroclinic term can be expressed in terms of the photospheric temperature
variation from equator to pole, $\Delta T$.  If this is to balance the
Coriolis term ${\cal C}$, then we must have relative temperature variations 
of the same order as in equation (\ref{eq:Svar}):
\begin{equation}\label{eq:tvar}
\frac{\Delta T}{T} \sim
\frac{2}{\pi} \frac{\pd}{\pd \theta} \ln \left<T\right> \sim 
\frac{2 R \lambda}{\pi g} \frac{\pd \Omega^2}{\pd z} \sim 2\times 10^{-5}  ~~.
\end{equation}
We have again assumed an ideal gas equation of state and hydrostatic
balance for simplicity.  If we further assume that the solar irradiance 
arises from blackbody emission, then we should expect to see a corresponding
latitudinal variation of the irradiance.  In other words, if the solar 
differential rotation in the NSSL were indeed in thermal wind balance 
[eq.\ (\ref{eq:tw})], then we would expect the poles to be brighter
than the equator with a relative irradiance enhancement of 
$\Delta I/I \sim 8 \times 10^{-4}$.  This corresponds to 
a temperature difference of about $0.1$K.

Irradiance variations of at least this order have indeed been detected
\citep[reviewed by][]{rast08}.  In particular, recent measurments by
\cite{rast08} indicate that the poles are indeed brighter than the 
equator, with $\Delta I/I \sim 1.5 \times 10^{-3}$.  However, whether
or not these can be interpreted as temperature variations is questionable.  
There is a strong possibility that the irradiance variation arises instead
from unresolved magnetic flux elements.  Thus, we view current estimates 
of ${\cal B}$ based on photospheric irradiance measurements as 
inconclusive as well.

The essential question is whether ${\cal B}$ can keep up with the
sharp increase in $\pd \Omega/\pd z$ throughout the NSSL.  The answer
depends on the details of convective heat transport in the solar
surface layers, which are not well understood.  If convection acts as
a turbulent thermal diffusion, then the thin shell geometry implies
that the latitudinal temperature gradient ($\Delta T$) at the base of the
NSSL, $r = r_s$, might be efficiently transmitted to the surface.  The
rapid decrease of the background temperature with radius might then
lead to an enhancement of the relative latitudinal temperature
gradient $\Delta T/T$, which in turn would imply an increase in 
${\cal B}$ [cf.\ eq.\ (\ref{eq:tvar})].  However, the horizontal 
heat transport and poleward meridional flow would tend to suppress 
latitudinal thermal gradients ($\Delta T$) and thus diminish ${\cal B}$.

To approach this issue more quantitatively, we estimate from helioseisimic 
inversions (\S\ref{sec:helio}) that ${\cal C}$ is a factor of 5-10 larger at 
the surface than at the base of the NSSL, $r = r_s \sim 0.946 R$ (defined
as where the mean radial $\Omega$ gradient changes sign; see
\S\ref{sec:homo}).  Over that same radial range the background
temperature decreases by a factor of 30-40.  Thus, taking into account
also the variation of $g$ and $r$, we estimate that horizontal
transport would have to be at least three times more efficient than
vertical transport in order for the magnitude of ${\cal B}$ to
increase less rapidly than ${\cal C}$, assuming thermal wind balance
at $r=r_s$ and that the photosphere is an isobaric surface, as in
equation (\ref{eq:tvar}).  We may expect the ratio of horizontal to 
vertical turbulent diffusion to scale as $\kappa_h/\kappa_v \sim U_h L_h / (U_v L_v)$
where $U$ and $L$ are typical velocity and length scales and $h$ and $v$
denote horizonal and vertical directions.  
Mass conservation, $\dv (\rho \vv) = 0$ implies $U_h/U_v \sim L_h/L_v$,
so $\kappa_h/\kappa_v \sim (L_h / L_v)^2$.  It is reasonable to expect
that this ratio may indeed exceed three for solar surface convection.
For example, if we put in numbers for granulation, $L_h \sim $ 1 Mm
and $L_v \sim H_\rho \sim$ 300 km, where $H_\rho$ is the density 
scale height, then $\kappa_h/\kappa_v \sim $ 10. This helps to substantiate
the conjecture that baroclinic forces may be too weak to balance the 
Coriolis force associated with the differential rotation in the NSSL.

In summary, we suggest that the characteristic change in the slope
of the $\Omega$ profile that defines the NSSL is largely determined 
by a shift in the meridional force balance from baroclinic
to turbulent stresses; that is, from equation (\ref{eq:tw}) to 
equation (\ref{eq:CG}).
Although this hypothesis is in principle testable, we believe that current
data is insufficient to either confirm or deny it.  In the remainder of
this paper we assume that this transition does indeed occur and we
explore its implications with regard to the dynamics of the NSSL.  

Note also that the statements made in this section apply mainly to the
persistent, background component of the mean flow that is present
throughout the solar cycle.  The baroclinic term ${\cal B}$ may well
establish time-varying meridional and zonal flows such as the
low-latitude branch of the solar torsional oscillation as discussed by
\cite{sprui03} and \cite{rempe07}.  In the Spruit-Rempel model,
enhanced cooling in magnetically active latitudinal bands induces
converging meridional flows that in turn accelerate zonal flows by
means of the Coriolis force.  If the cooling is sustained for at least
several months (long enough for the Coriolis force to fully respond),
then the dynamical balances in equation (\ref{eq:gp}) and
(\ref{eq:tw}) can be established in a quasi-static sense.  As for the
persistent background flow, the coupled inertia of the fluctuating
flow components provides a tight link between the meridional
circulation and the rotational shear.

\section{Helioseismic Estimates of Turbulent Transport}\label{sec:helio}

In \S\ref{sec:gp} and \S\ref{sec:meridional} we argued that the dynamical 
balances that are likely to prevail in the NSSL are given by 
equations (\ref{eq:gp}) and (\ref{eq:CG}).  The left-hand side of each
of these equations involves quantities that are accessible to helioseismic
inversions, namely $\Omega$ and $\left<\vv_m\right>$.  Thus, if we 
assume these balances hold, then we can obtain observational estimates 
for the turbulent stresses ${\cal F}$ and ${\cal G}$.

That is the focus of this section.  We will obtain estimates for the
turbulent stresses ${\cal F}$ and ${\cal G}$ based on local and
global helioseismic inversions and we will discuss their 
implications for the nature of turbulent transport in the NSSL.
Then, in \S\ref{sec:transport} we will investigate whether these
results are consistent with simple theoretical models.  We begin
in \S\ref{sec:data} by discussing the data we use and we then
proceed to calculate ${\cal F}$ and ${\cal G}$ in \S\ref{sec:fghelio}.

\subsection{Helioseismic Data and Inversions}\label{sec:data}

We use both local and global helioseismic procedures to estimate the
solar rotation rate $\Omega$ and the meridional flow component
$\left<{\bf v}_m\right>$. Both techniques exploit the fact that,
through the Doppler effect, a flow induces a frequency splitting between
acoustic waves propagating in opposite directions. By measuring this
frequency splitting for a variety of waves that reside within different
regions of the solar interior, through an inversion process, maps of the
flow as a function of latitude and depth can be produced. Global
helioseismology employs modes of long horizontal wavelength that
comprise the sun's global acoustic resonances. The frequency splittings
of such modes can be used to measure the axisymmetric component of the
rotation rate over a broad range of latitudes (within $70^\circ$ of the
equator) and to depths spanning the entire convection zone. Local
helioseismic techniques, which measure short-wavelength waves, do not
permit sampling of the flows to such high latitudes or to such large
depths (only within $50^\circ$ of the equator and only within the upper
15 Mm or 2\% by fractional radius); however, local helioseismology is
capable of measuring both the rotation rate and the meridional component
of the flow, unlike global helioseismology.

We utilize a local helioseismic procedure called ring analysis which
assesses the speed and direction of subsurface horizontal
flows by measuring the advection of ambient acoustic waves by those
flows \citep[e.g.][]{hill88,haber02}. Measurement of the frequency
splitting provides a direct measure of the fluid's flow velocity in
those layers where the waves have significant amplitude. The frequency
shift induced by the flow on any given wave component is given by
$\Delta\omega = {\bf k} \bdot {\bf \bar{U}}$, where ${\bf k}$ is the
wave's horizontal wavenumber and ${\bf \bar{U}}$ is the integral over
depth of the horizontal flow velocity weighted by a kernel which is
approximately the kinetic energy density of the acoustic wave. The
frequency splittings for a large number of waves of different wavenumber
and radial mode order form a system of integral equations that can be
inverted to obtain the horizontal flow as a function of depth 
\citep{thomp96,haber04}.  By repeating the analysis over many
different locations on the solar surface and over many different days, a
map of the horizontal flow as a function of longitude, latitude, depth
and time can be generated.

\begin{figure}
\centerline{\epsfig{file=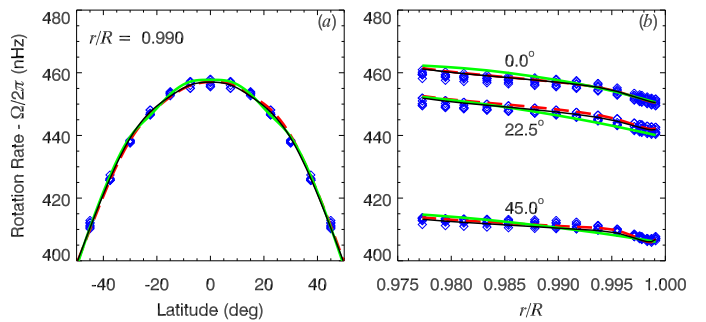,width=3.5in}}
\caption{Rotation profile $\Omega$ obtained from local and
global inversions, plotted verses ($a$) latitude (at the indicated radius) 
and ($b$) radius (at the indicated latitudes).  Inversions
for the year 1996 are shown as a black line and other years
1997-2003 are shown as blue symbols.  The green
curve represents the global rotation inversions shown
in Figure \ref{fig:DR} and the dashed red curve represents
an analytic fit to the local 1996 inversions of the form expressed
in equation (\ref{eq:omega}) below.\label{fig:omega_helio}}
\end{figure}

\begin{figure}
\centerline{\epsfig{file=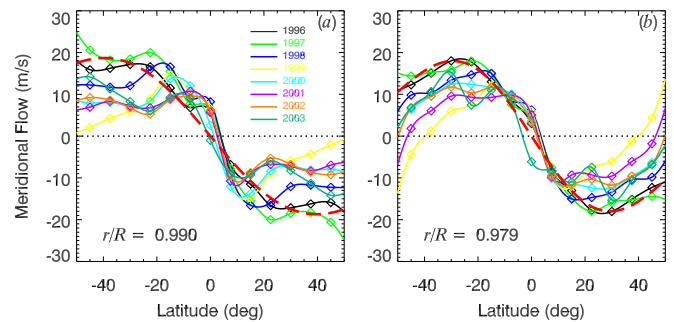,width=3.5in}}
\caption{Longitudinally-averaged (co-) latitudinal velocity $\left<v_\theta\right>$
obtained from local helioseismic inversions (ring analysis), plotted versus
latitude at ($a$) $r = 0.990 R$ and ($b$) $r = 0.979 R$.  Black lines
and symbols denote data from 1996 and other colors denote other years,
as indicated.  We focus on 1996 but we show other years merely to highlight 
the magnitude of the variations; the relationship between meridional flow
variations and solar activity has been investigated in detail elsewhere
\citep[e.g.][]{hatha96b,haber02,zhao04,gonza06,ulric10,hatha10,hatha11}.
The red dashed lines represent a fit to the 1996 inversions as expressed
in equation (\ref{eq:vtfit}).\label{fig:vth_helio}}
\end{figure}

Our local helioseismic inferences were made using Dopplergram data from
the Michelson Doppler Imager (MDI) aboard the Solar and Heliospheric
Observer (SoHO). Flows have been obtained using the Dynamics Campaign
data, of which 2-3 months of data per year is available. We have
chosen to use data from the rising phase of the solar cycle from 1996
through 2003. The short-wavelength waves used by local helioseismic
procedures are trapped near the solar surface, and the ring-analysis
measurements used here are therefore capable of sampling down to a depth
of 15 Mm below the photosphere, or the upper half of the NSSL. 
Each local analysis samples a region of Sun that is roughly 180 Mm
(15$^\circ$ in heliographic angle) in diameter.  On any given day,
a mosaic of 189 overlapping analysis regions are 
independently analyzed, resulting in a flow map with a horizontal
spacing of 7.5$^\circ$ between measurement regions, each with a
horizontal resolution of 15$^\circ$.  

Estimates
for the mean rotation rate $\Omega$ and the latitudinal component of the
meridional flow $\left<v_\theta\right>$, have been obtained by averaging the
resultant maps over all longitudes and over the 2-3 months that the
Dynamics Campaign data is available each year. The mean rotation rate
$\Omega$ is illustrated in Figure \ref{fig:omega_helio}. The black curve 
corresponds to measurements made in 1996, while the blue diamonds show the 
weak variability in the flows during the years 1997 to 2003. The latitudinal
component of the flow, $\left<v_\theta\right>$, is shown in 
Figure \ref{fig:vth_helio} at two different depths and with 
a differently-colored curve for each year.

The global helioseismic assessments were made using four non-overlapping
60 day intervals of GONG data from the year 1996 \citep{howe00,schou02}.
The rotation rate is inferred from the frequency
splitting between modes of equal but opposite azimuthal order $\pm m$.
Such modes are identical in all respects other than the fact that they
propagate in opposite directions in longitude around the sun. Solar
rotation Doppler shifts the mode frequencies in opposite directions and
the frequency splittings of many different modes can be inverted to
obtain the rotation rate as a function of latitude and depth. The
result of a regularized Least Squares (RLS) inversion is shown 
in Figures \ref{fig:DR} and \ref{fig:omega_helio}.

\subsection{Estimating ${\cal F}$ and ${\cal G}$}\label{sec:fghelio}

If we assume the dynamical balances in equations (\ref{eq:gp}) and (\ref{eq:CG})
hold, then we can turn them around and calculate ${\cal F}$ and ${\cal G}$
based on helioseismic inversions.  In particular, if we assume that to lowest
order $\rho \approx \left<\rho\right>$, then equation (\ref{eq:gp}) yields
an estimate of the specific torque
\begin{equation}\label{eq:Fhelio}
{\cal T} = \frac{{\cal F}}{\rho} = \left<\vv_m\right> \bdot \del {\cal L}
= \left<v_r\right> \frac{\pd {\cal L}}{\pd r} + \frac{\left<v_\theta\right>}{r} 
\frac{\pd {\cal L}}{\pd \theta}  ~~~.
\end{equation}
Thus, the right-hand-side only depends on ${\cal L} = \lambda \left<v_\phi\right> = \lambda^2 \Omega$ 
and $\left<\vv_m\right>$.  Similarly, equation (\ref{eq:CG}) gives
\begin{equation}\label{eq:Ghelio}
{\cal G} = - \lambda \frac{\pd \Omega^2}{\pd z}
\end{equation}
which only depends on $\Omega$.

\begin{figure}
\centerline{\epsfig{file=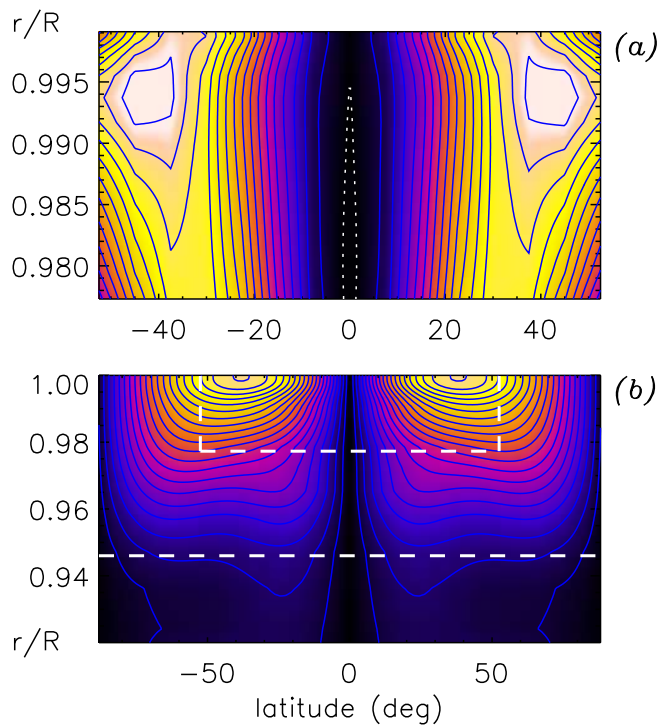,width=3.5in}}
\caption{Estimates for the turbulent stresses ($a$) ${\cal T} = {\cal F} \rho^{-1}$
and ($b$) ${\cal G}$, computed from equations (\ref{eq:Fhelio}) and (\ref{eq:Ghelio}) 
using local and global helioseismic inversions from 1996, as described in the text.  
In ($a$), the color table ranges from
-3.95$\times 10^8$ cm$^2$ s$^{-1}$ (white) to zero (black).  Blue contours denote
negative values and the white dotted line indicates the zero contour.  In ($b$)  
the color scale runs from zero (black) to 1.28$\times 10^{-11}$ s$^{-2}$, with
blue contours denoting positive values.  Note that the vertical and horizontal
axes are different in frames ($a$) and ($b$) due to the limited extent of
the local inversions.  The region spanned in ($a$) is indicated in ($b$) by
a white dashed line.  A second dashed line indicates the base of the 
NSSL, $r_s = 0.946 R$, defined as where the spherically-averaged $\Omega$
gradient changes sign.\label{fig:FG}}
\end{figure}

Estimates for ${\cal F}$ and ${\cal G}$ based on equations (\ref{eq:Fhelio}) and 
(\ref{eq:Ghelio}) are shown in Figure \ref{fig:FG}.  In the remainder of this section
we provide further details on how these results were obtained.  For both ${\cal F}$ 
and ${\cal G}$ we focus on data from 1996.  Since this corresponds
to solar minimum, it minimizes the influence of the cyclic component of the solar
magnetic field, focusing instead on the persistent turbulent stresses that maintain
the NSSL throughout the solar cycle (\S\ref{sec:variability}).

Before proceeding to the more involved calculation of ${\cal F}$, we first consider 
the estimate for ${\cal G}$ shown in Figure \ref{fig:FG}$b$.  Since this only 
depends on $\Omega$, we use the global inversions shown in Figure \ref{fig:DR} 
to compute it.  These extend to higher latitudes and lower radii than the local inversions
and the longitudinal coverage provided by global modes may more accurately reflect the 
axisymmetric $\Omega$ gradients in the NSSL.  In order to compute the latitudinal
$\Omega$ gradient, we first fit the $\Omega$ profile to a functional form given by
\begin{equation}\label{eq:omega}
\Omega(r,\theta) = \Omega_e(r) + \Omega_2(r) \cos^2\theta + \Omega_4(r) \cos^4\theta
\end{equation}
and then compute the $\theta$ gradient analytically.  We then compute the 
radial $\Omega$ gradient by means of a second-order finite difference scheme 
(on a non-uniform grid) applied to the fitting coefficients $\Omega_e$, 
$\Omega_2$, and $\Omega_4$.  

The estimation of ${\cal T}$ from equation (\ref{eq:Ghelio}) involves correlations
between the mean meridional and zonal flow.  For these we use the local inversions
described in \S\ref{sec:data}.  In particular, we fit the local $\Omega$ inversion
to a functional form as in equation (\ref{eq:omega}) and the local $\left<v_\theta\right>$
inversion (both from 1996) to a functional form given by
\begin{equation}\label{eq:vtfit}
\left<v_\theta\right> = \sin\theta \cos\theta \left(c_1(r) + c_3(r) \cos^2\theta\right)  ~~~.
\end{equation}
The fits to the local $\Omega$ and $\left<v_\theta\right>$ inversions are plotted 
in Figures \ref{fig:vth_helio} and \ref{fig:omega_helio} as red dashed lines.
The global $\Omega$ fit lies almost on top of these other curves and is 
thus omitted for clarity.

\begin{figure}
\centerline{\epsfig{file=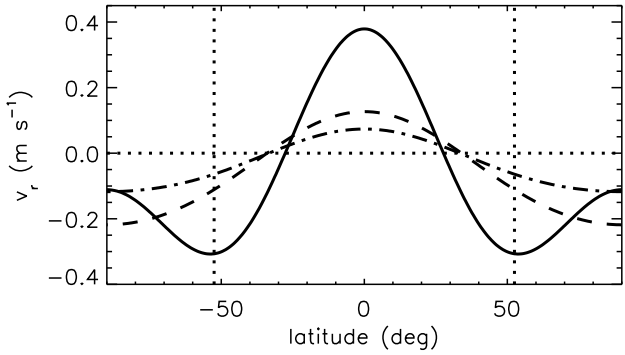,width=3.5in}}
\caption{Mean radial velocity $\left<v_r\right>$ inferred from mass 
conservation, based on local helioseismic inversions for $\left<v_\theta\right>$.
Solid, dashed, and dot-dashed lines correspond to $r = $ 0.977$R$,
0.990$R$, and 0.996$R$ respectively.  Dotted lines indicate zero velocity
and the range in latitude spanned by the local inversions ($\pm$ 52.5$^\circ$).
Note that the curves extend beyond this range because they are based on 
the analytic expression given in equation (\ref{eq:vrex}).\label{fig:vr}}
\end{figure}

The mean radial velocity $\left<v_r\right>$ is obtained by assuming mass conservation;
$\dv \left<\rho \vv_m\right> = 0$.  We also assume that
$\rho \approx \left<\rho\right> \approx \rho_s(r)$ where $\rho_s(r)$ is 
the spherically-symmetric mean density given by solar structure Model S
of \cite{chris96}.  
This gives
\begin{equation}\label{eq:vreq}
\frac{\pd }{\pd r} \left(r^2 \rho_s \left<v_r\right>\right) = 
\frac{r \rho_s}{\sin\theta} \frac{\pd }{\pd \theta} \left(\sin\theta \left<v_\theta\right>\right)  ~~~.
\end{equation}
The $\theta$ derivative on the right-hand-side of equation (\ref{eq:vreq}) is computed analytically
from the fit in equation (\ref{eq:vtfit}).  This leads to the following form for $\left<v_r\right>$
\begin{equation}\label{eq:vrex}
\left<v_r\right> = w_e(r) + w_2(r) \cos^2\theta + w_4(r) \cos^4\theta ~~~. 
\end{equation}
We then proceed by expressing the radial derivative on the right-hand-side 
of equation (\ref{eq:vreq}) by means of a second-order finite difference scheme and we 
solve the resulting matrix equations for $r^2 \rho_s w_i$ ($i$ = $e$, 2, 4), with boundary 
conditions such that $\left<v_r\right> = 0$ at $r = R$.  The $w_i$ are then smoothed using 
a boxcar window, taking into account the non-uniform grid.  The 
resulting $\left<v_r\right>$ profile is shown in Figure \ref{fig:vr} for several radii.

Note that the amplitude of the inferred radial velocity is extremely small 
in the NSSL, less than 1 m s$^{-1}$ and its contribution to ${\cal F}$ 
(and ${\cal T}$) is negligible.  It does contribute a small positive
maximum at the equator ($\sim 0.07 \times 10^8$ cm$^2$ s$^{-1}$) 
as seen in Figure \ref{fig:FG}$a$, but the prominent maximum
at mid-latitudes is due almost entirely to the final term in
equation (\ref{eq:Fhelio}), proportional to 
$\left<v_\theta\right> \pd {\cal L}/\pd \theta$.  
Although there is some uncertainty in the estimation
of $\left<v_r\right>$ (\S\ref{sec:variability}), we believe this conclusion is robust.
In order for the radial component to make a significant
contribution to ${\cal F}$, the radial velocity would have to 
be at least an order of magnitude larger than our estimate.
This would imply either a latitudinal velocity that is an order 
of magnitude larger, of order 200 m s$^{-1}$, or a latitudinal
gradient of $\left<v_\theta\right>$ that is an order of magnitude 
larger.  If the amplitude of the meridional flow is to be no 
larger than 20 m s$^{-1}$, the requisite gradient would require 
that the peak flow be achieved within 3 degrees of the equator.
Both scenarios can be ruled out by surface observations
and helioseismic inversions.

Both ${\cal F}$ and ${\cal G}$ peak at mid-latitudes (Fig.\ \ref{fig:FG})
because this is where 
$\pd \Omega/\pd z$ and $\left<v_\theta\right> \pd {\cal L}/\pd \theta$
peak.  In fact, these two terms ($\pd \Omega/\pd z$ and 
$\left<v_\theta\right> \pd {\cal L}/\pd \theta$) are identically zero 
at the equator because of the symmetry implied by equations 
(\ref{eq:omega}) and (\ref{eq:vtfit}).  As emphasized in 
\S\ref{sec:anisotropy}, this does not mean that the local
angular and meridional momentum flux must vanish at the equator, nor
must the mean shear.  Rather, it implies that the turbulent transport 
and the mean flows adjust themselves in such a way as to minimize the 
net zonal flux divergence (axial torque) ${\cal F}$ and the net
meridional ``force curl'' ${\cal G}$.  This is in
contrast to mid-latitudes, where the net zonal and meridional momentum
transport represented by ${\cal F}$ and ${\cal G}$ must be nonzero 
in order to balance the inertial forces associated with the mean 
differential rotation and meridional circulation.  This delicate, 
nonlinear, nonlocal interplay between turbulent transport and 
mean flows ultimately determines the structure of the NSSL.

Note that the sign of ${\cal F}$ (and ${\cal T}$) is negative
at mid-latitudes, signifying that turbulent stresses must exert 
a retrograde net axial torque, removing the angular momentum that 
is imparted by the meridional flow.  The sign of ${\cal G}$, 
meanwhile, is positive, tending to induce a clockwise circulation
in the northern hemisphere in order to offset the 
counter-clockwise circulation that would tend to establish
a cylindrical (Taylor-Proudman) rotation profile (\S\ref{sec:thex}).

\subsection{Anisotropy and Energetics}\label{sec:anen}

The quantities plotted in Figure \ref{fig:FG}, ${\cal T} = {\cal F} \rho^{-1}$
and ${\cal G}$, represent turbulent (non-axisymmetric) zonal and meridional
momentum transport but a direct comparison between them is somewhat elusive 
since they apply in different contexts; ${\cal F}$ appears in the angular momentum 
equation while ${\cal G}$ appears in the zonal vorticity equation; their units are 
different.  In this section we put ${\cal F}$ and ${\cal G}$ on an equal
footing by converting them to accelerations, with units of cm s$^{-2}$.
In other words, we can write the momentum equation as
\begin{equation}\label{eq:mom}
\frac{\pd \left<\vv\right>}{\pd t} = - \left(\left<\vv\right> \bdot \del\right) \left<\vv\right> 
+ \frac{\del \left<P\right>}{\left<\rho\right>} + {\bf A} ~~~.
\end{equation}
Note that equation (\ref{eq:mom}) is exact in the sense that it is faithful
to the compressible MHD equations, with all non-axisymmetric influences
(as well as some axisymmetric influences such as viscous diffusion and
the mean-field Lorentz force) encapsulated in the ``turbulent'' 
acceleration term ${\bf A}$.

Now consider again our dynamical balance equations (\ref{eq:gp}) and 
(\ref{eq:CG}).  The left-hand-side of both of these equations arises
from the first term on the right-hand-side of equation (\ref{eq:mom}),
namely the inertia associated with mean flows (we use an inertial 
reference frame so the Coriolis force is included in this term
as well; see Appx.\ A).  The terms on the right-hand-side of 
equations (\ref{eq:gp}) and (\ref{eq:CG}), ${\cal F}$ and 
${\cal G}$, arise from ${\bf A}$ so we
can compute ${\bf A}$ from the helioseismic results presented in
Figure \ref{fig:FG}.  Note that the second term on the right-hand-side
of equation (\ref{eq:mom}) gives rise to the baroclinic term ${\cal B}$ in equations
(\ref{eq:mer}) and (\ref{eq:tw}), which we propose is negligible in the NSSL
(\S\ref{sec:what}).

\begin{figure}
\centerline{\epsfig{file=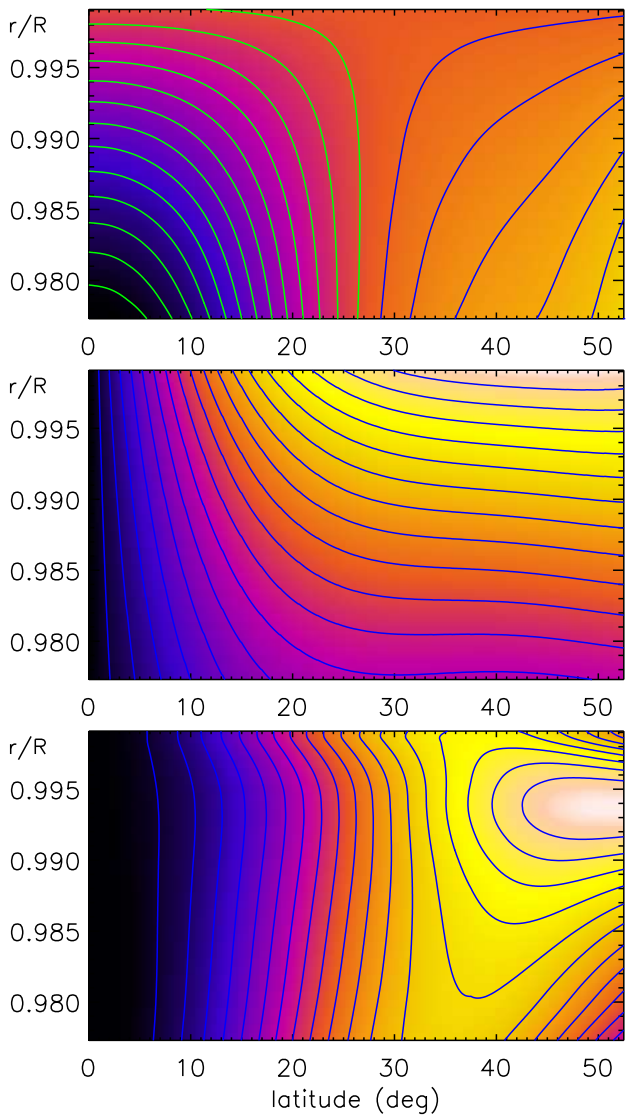,width=3.5in}}
\caption{Acceleration terms (top) $A_r$, (middle) $A_\theta$, 
and (bottom) $- A_\phi$, for the region spanned by the local 
inversions ($r > 0.978 R$, lat $\leq 52.5^\circ$). Scale ranges
are $\pm 9\times 10^{-4}$ cm s$^{-2}$, 0--$2\times 10^{-2}$ cm s$^{-2}$, 
and 0--$8\times 10^{-3}$ cm s$^{-2}$ respectively, with colors ranging
from black/blue to yellow/white.  Blue contours denote positive values
and green negative.
\label{fig:gyan2d}}
\end{figure}

\begin{figure*}
\centerline{\epsfig{file=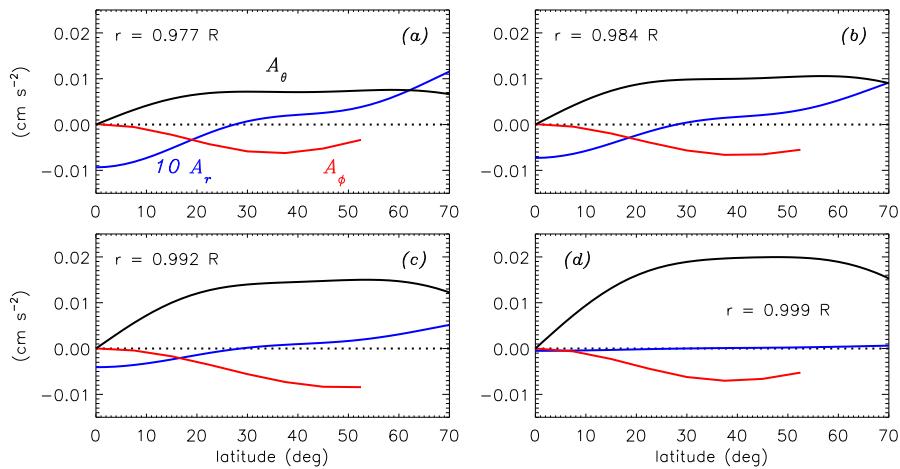,width=5in}}
\caption{Acceleration terms $A_r$ (dashed line, blue in online
version), $A_\theta$ (solid black line), and $A_\phi$ (dot-dashed line;
red in online version) as in Figure \ref{fig:gyan2d} but here plotted 
together versus latitude for several different radii, as indicated.
The radial component $A_r$ is multiplied by a factor of 10 
for clarity of presentation.  Note that we only show the $A_\phi$ 
curve out to 52.5$^\circ$ latitude because that is the extent of
the local inversions used to compute it.\label{fig:gyan}}
\end{figure*}

The zonal component of ${\bf A}$ is directly related to ${\cal F}$:
\begin{equation}\label{eq:Aphi}
A_\phi = \frac{{\cal F}}{\lambda \rho}  ~~~.
\end{equation}
In order to obtain a relationship between ${\cal G}$ and the meridional
components of ${\bf A}$, ${\bf A}_m$, we note that any arbitrary,
axisymmetric vector field with only meridional components can be
represented as follows:
\begin{equation}
{\bf A}_m = \curl \left(\zeta \uvp\right) + \del \chi ~~~,
\end{equation}
where $\zeta$ and $\chi$ are functions of $r$ and $\theta$.
The helioseismic estimate of ${\cal G}$ is obtained from the zonal vorticity
equation and therefore provides no information about the compressible 
component, $\chi$.  However, taking the curl of equation (\ref{eq:mom})
yields an equation for $\zeta$:
\begin{equation}\label{eq:zetaeq}
\left(\curl {\bf A_m}\right) \bdot \uvp = 
- \nabla^2 \zeta + \frac{\zeta}{\lambda^2}  = {\cal G} ~~~.
\end{equation}
We solve this equation as described in Appendix C, subject to boundary
conditions such that the radial acceleration vanish at the surface
($A_r = 0$ at $r = R$), and the latitudinal acceleration vanish at
the base of the NSSL ($A_\theta = 0$ at $r = r_s$).  The latter
condition implies that the latitudinal acceleration is relative 
to the deep CZ.  The results are shown in Figures \ref{fig:gyan2d}
and \ref{fig:gyan}.

To facilitate comparison between the terms, the plotting range in
Figure \ref{fig:gyan2d} is limited to that spanned by the local
inversions.  However, the results for $A_r$ and $A_\theta$ 
extend to higher latitudes and deeper levels because they
are based on the global inversions in Figure \ref{fig:DR}.
This greater latitudinal extent is reflected in Figure 
\ref{fig:gyan}, which also serves to illustrate the relative
amplitudes and signs of each term.

The amplitudes of $A_\theta$ and $A_\phi$ peak at mid-latitudes
while $A_r$ is radially inward at the equator and outward above
latitudes of about 26$^\circ$.  As noted above with regard
to Figure \ref{fig:FG}$a$, the zonal acceleration is 
negative, tending to decelerate the rotation rate in the
NSSL.  The positive sign of $A_\theta$ indicates an equatorward
direction, tending to oppose the poleward meridional flow,
thereby maintaining the rotational shear against Coriolis-induced
circulations.  The reversing sign of $A_r$, with an inward
orientation at the equator also tends to oppose the meridional
flow.  This is consistent with the conceptual framework put 
forth in \S\ref{sec:gp} and \S\ref{sec:meridional}.  

The amplitudes of $A_\theta$ and $A_\phi$ are comparable,
which is remarkable because they are approximated by
means of very different data sets and analysis techniques
and represent distinct dynamical balances.  The amplitude 
of the vertical acceleration is roughly an order of magnitude 
smaller than the horizontal components.

We can estimate the total rate of work done by each of the turbulent
stress components as follows:
\begin{equation}
W_i = \int_V \rho_s \left<v_i\right> A_i dV  
\mbox{\hspace{.5in} ($i = r, ~ \theta, ~ \phi$)}
\end{equation}
where $V$ is the volume of the NSSL and $\rho_s$ is again the
spherically-symmetric density given by Model S.  The results
are listed in Table 1.

In the first column of Table 1, the volume $V$ corresponds to
the region spanned by the local helioseismic inversions and
plotted in Figure \ref{fig:gyan2d}.  In the second column
we extrapolate this to the entire NSSL, spanning
$0.95 R$--$1.00 R$ in radius and pole-to-pole in latitude.
The poleward extrapolations are based on the analytic fits 
expressed in equations (\ref{eq:omega}) and (\ref{eq:vtfit}),
using global inversions for $\Omega$ and local inversions
for $\left<v_\theta\right>$.  

\begin{figure}
\centerline{\epsfig{file=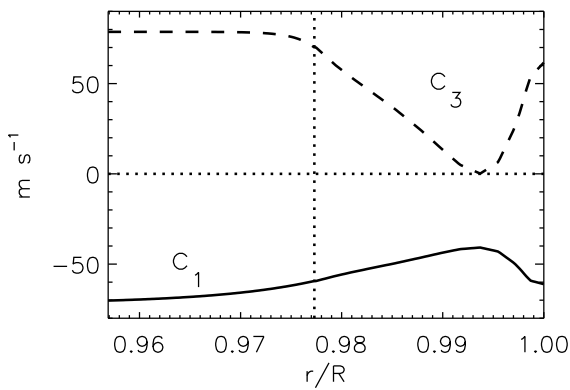,width=3.0in}}
\caption{Radial variation of the expansion coefficients
$c_1(r)$ (solid line) and $c_3(r)$ (dashed line), defined 
in equation (\ref{eq:vtfit}).  For $r > 0.977R$ (indicated
by the vertical dotted line), the curves are based on fits
to local helioseismic inversions.  The extrapolations for
$r < 0.977 R$ are chosen to have continuous values and
derivatives but flatter slopes as they extend deeper.
\label{fig:cex}}
\end{figure}

\begin{deluxetable}{ccc}
\tablecaption{Energetics and Time Scales\tablenotemark{a}}
\tablehead{ & \colhead{$r > 0.978 R$, lat $< 52.5^\circ$} & 
\colhead{$r > 0.95 R$ (extrapolated)}}
\startdata
$W_r$ & -8.7$\times 10^{-8} L$ & -2.5$\times 10^{-6} L$ \\
$W_\theta$ & -1.3$\times 10^{-4} L$ & -4.7$\times 10^{-4} L$ \\
$W_\phi$ & -1.9$\times 10^{-2} L$ & -4.2$\times 10^{-2} L$ \\
$W_g$ & 8.9$\times 10^{-4} L$ & 6.7$\times 10^{-3} L$ \\
$W_t$ & -2.0$\times 10^{-2} L$ & -4.9$\times 10^{-2} L$ \\
$\tau_s$ & 200 days & 590 days 
\enddata
\tablenotetext{a}{Work values are given in terms of the solar luminosity $L$.}
\end{deluxetable}

When extrapolating $A_\phi$ in particular, we compute $\del {\cal L}$
in equation (\ref{eq:Fhelio}) from global inversions interpolated
onto the local helioseismic grid for $r \geq 0.977 R$.  We then 
extrapolate the expansion coefficients $c_1(r)$ and $c_3(r)$ that 
appear in (\ref{eq:vtfit}), as shown in Figure \ref{fig:cex}, 
in order to obtain $\left<v_\theta\right>$ and $\left<v_r\right>$
down to $0.95R$.  Slopes are chosen to flatten out below the matching 
point at $r = 0.977 R$, anticipating an eventual flow reversal
deeper in the CZ (we choose $c_3$ to flatten out somewhat
more rapidly than $c_1$ in order to avoid high-latitude 
counter-cells resulting solely from the extrapolation).  
The numbers in Table 1 are not expected to be greatly 
sensitive to uncertainties in the meridional flow structure, 
unless the flow reversal actually occurs within
the NSSL (\S\ref{sec:variability}).

The values in Table 1 indicate that the zonal force $A_\phi$
(corresponding to the net axial torque ${\cal F}$) does the most
work, accounting for 2-4\% of the solar luminosity.  The
work done by the latitudinal force $A_\theta$ is roughly two 
orders of magnitude smaller and that done by the radial
force $A_r$ is two to three orders of magnitude smaller 
still.

If the net axial torque ${\cal F}$ can be represented as the
divergence of an angular momentum flux ${\cal F} = - \dv {\bf F}$,
then the zonal work can be expressed as
\begin{equation}
W_\phi = \int_V {\bf F} \bdot \del \Omega dV - \int_{\cal S} \Omega {\bf F} \bdot d\surf
\equiv W_g + W_t
\end{equation}
where ${\cal S}$ is the bounding surface of $V$, and $d\surf$ is directed normal to 
the surface.  Here $W_g$ is the work done
against the angular velocity gradient and $W_t$ is the work done by means of 
transporting angular momentum through the surface ${\cal S}$.  In a stationary state 
the mean mass flux is divergenceless, so equation (\ref{eq:gp}) implies
${\bf F} = - \left<\rho \vv_m\right> {\cal L}$.  The resulting values 
of $W_g$ and $W_t$ are listed in Table 1 and indicate that the latter 
dominates $W_\phi$, accounting for its negative sign.  Thus, most of 
the work done by turbulent stresses to maintain the NSSL against 
advection by the meridional flow involves transporting angular 
momentum out of the layer.

Our estimate for the angular momentum flux, ${\bf F}$, also allows
us to compute a spin-down time scale for the NSSL:
\begin{equation}
\tau_s = \frac{\int_V \rho {\cal L} dV}{\int {\bf F} \bdot d{\bf S}}  ~~~.
\end{equation}
As indicated in Table 1, this suggest a value of about 200 days for the
region spanned by the local inversions and about 590 days for the
entire NSSL.  This is comparable to the ventilation time
\begin{equation}
\tau_v \sim \frac{\pi R}{2 V_\theta} \sim \mbox{630 days ,}
\end{equation}
where $V_\theta \sim $ 20 m s$^{-1}$ is a typical meridional
flow amplitude. This correspondence is not surprising since we estimated 
the turbulent angular momentum flux ${\bf F}$ based on the meridional flow.

\subsection{Variability and Uncertainty}\label{sec:variability}

As mentioned in \S\ref{sec:intro} and as demonstrated in \S\ref{sec:data} 
(Fig.\ \ref{fig:vth_helio}), estimates of the meridional flow based on different 
observational data sets and different analysis techniques can vary substantially.  
Some of this variability can be attributed to measurement uncertainities but some 
is undoubtedly due to the intrinsic variability of the flow itself.  In addition 
to the meridional flow, equations (\ref{eq:gp}) and (\ref{eq:mer}) also involve
gradients in the angular velocity $\Omega$ which are also subject to measurement
uncertainties and intrinsic variability.  In this section we address what implications
this variability and uncertainty has with regard to the principle arguments, results,
and conclusions presented in this paper.

We begin with the physical processes discussed in sections 
\ref{sec:gp} and \ref{sec:meridional}.  The predominant dynamical
balances we advocate for the NSSL, represented by equations
(\ref{eq:gp}) and (\ref{eq:CG}), are expected to hold regardless
of the detailed amplitude and structure of the mean flows.
They assume only that well-defined, persistent zonal and meridional
flows exist when averaged over time intervals longer than a rotation
period (of order one month) and that turbulent (i.e.\ Reynolds)
stresses dominate over baroclinic forces (${\cal G} >> {\cal B}$)
in the NSSL.

Global rotational inversions indicate that variations in $\Omega$ over
the solar cycle are not more than a few percent \cite{thomp03}.
Although $\Omega$ gradients may vary more than this (see below), they
are not large enough to reverse the sign of $\pd {\cal L}/\pd
\lambda$.  In order to do so, relative variations in the rotation rate
$\Delta \Omega/\Omega$ would have to exceed $2 D/R$ where $D$ is the
length scale of the shear.  Taking $D$ to be the extent of the NSSL
$\sim 0.05 R$ implies variations with time of the zonal flow of at
least several hundred m s$^{-1}$.  Such large temporal variations can
be ruled out based on photospheric observations and helioseismic
rotational inversions (torsional oscillations are more like 20 m
s$^{-1}$).  Likewise, the uncertainty in $\rho$ (obtained from Model
S) is constrained by helioseismic structure inversions and is thus
certain to within a few percent.

Thus, the principle source of uncertainty in the quantitative
estimates presented in sections \ref{sec:fghelio} and \ref{sec:anen}
is the meridional flow.  Relative to zonal flow inversions, meridional
flow inversions are subject to both greater measurement errors and
greater intrinsic variability, by virtue of their weaker amplitude and
asymmetric structure with respect to the equator (which precludes
global inversions).  Note that this source of uncertainty only effects our
estimate for ${\cal F}$ and related quantities such as ${\cal T}$ in
Fig.\ \ref{fig:FG}$a$, $A_\phi$ in Figures \ref{fig:gyan2d}$c$ and
\ref{fig:gyan}, and $W_\phi$, $W_g$, $W_t$, and $\tau_s$ in Table 1.
Our estimate for ${\cal G}$ and related quantities ($A_r$, $A_\theta$,
$W_r$, $W_\theta$) depends only on global rotation inversions, which
are generally more reliable, particularly at high latitudes and large
depths below the photosphere.

As noted in \S\ref{sec:fghelio}, the radial flow $\left<v_r\right>$
makes a negligible contribution to our estimate of ${\cal F}$, 
even when the variability and uncertainty in the meridional
inversion is taken into account.  We have verified that this
is consistent with our data by computing ${\cal T}$ for each
of the years shown in Figures \ref{fig:omega_helio} and
\ref{fig:vth_helio} (1996-2004).  The amplitude of the radial 
component never exceeds 2$\times 10^7$ cm$^2$ s$^{-1}$, which 
is 5\% of the saturation value of the color scale used in 
Figure \ref{fig:FG}$a$.   The average amplitude is less than 
one percent of this saturation value. 

The time variation of our $\left<v_\theta\right>$ inversions varies greatly
with latitude and depth, as shown in Figure \ref{fig:vth_helio}. 
At latitudes less than 40$^\circ$ the standard deviation of these
measurements is less than 50\%.  This is also the case for our estimate
of ${\cal T}$, confirming that $\left<v_\theta\right>$ is the principle
source of uncertainty and variation.  At latitudes greater than 40$^\circ$
and radii below 0.99$R$ the inversions become less reliable and the standard
deviation exceeds 50\%. 

However, we remind the reader that our quantitative estimates for
${\cal F}$ and related quantities in \ref{sec:fghelio} and
\ref{sec:anen} are based on an analytic fit to the 1996 inversions, as
shown by the dashed line in \ref{fig:vth_helio} (1996-2004).  This is
a smooth poleward flow with an amplitude of 18-21 m s$^{-1}$ at all
measured radii $r \geq 0.977 R$.  Although the amplitude of this flow
is debatable, this general profile is consistent with the inpretation
put forth by several authors who have argued that the spatial and
temporal variation of the meridional flow can be interpreted as a
persistent poleward background flow plus a time-dependent component
associated with cyclic magnetic activity
\cite[e.g.][]{snodg96,basu10,hatha11}.  We are interested mainly in the
background flow which is present throughout the solar cycle and is
most prominent during solar minimum.

We emphasize that time averages of at least 2-3 months are needed
to properly assess the mean flows we are concerned with here.
That is the time scale over which the dynamical balances in 
equations (\ref{eq:gp}) and (\ref{eq:CG}) are established
by means of the Coriolis force.  Estimates of the meridional 
flow vary significantly from one Carrington rotation to the next,
independent of the measurement technique. For example, the results
of Hathaway \& Rightmire (2010) based on feature tracking indicate
that the amplitude of the dominant Legendre component can change by
as much as 5 m s$^{-1}$ over the course of one year. If one includes 
the higher-order components, the variation can be even larger, of 
order 5-10 m s$^{-1}$ at the higher latitudes.

The importance of temporal averaging is evident when comparing our
meridional flow inversions to those of Basu \& Antia (2010), which are
based on the same MDI data sets but each averaged over a single
Carrington rotation.  The natural variability in our measurements
should be reduced from those of Basu \& Antia by roughly a factor of
1.5 just due to our longer averages.  Furthermore, our estimates are
obtained by averaging over both longitude and time. At any given
instant in time, we make 189 separate flow determinations scattered
across the solar disk. All measurements made at the same latitude are
then averaged together (15 determinations at the equator and 7 at the
highest latitudes). The process is repeated on a daily basis
and the results further averaged in time. Basu \& Antia, on the other
hand, only make estimates along the central meridian. So at high
latitudes, even if the two teams average over a similar duration, we
have 7 times more data going into the average. This high degree of
temporal variability is the reason we used a representative smooth fit
to the helioseismic measurements instead of directly using the
measurements themselves.

According to our inversions, the amplitude of the background
poleward flow is approximately 20 m s$^{-1}$.  Other estimates
based on various techniques including heliseismic inversions,
Doppler measurements, and feature tracking generally range
from 10-20 m s$^{-1}$
\citep{snodg96,hatha96b,haber02,zhao04,hindm04,gonza06,ulric10,basu10,hatha10,hatha11}.
If the mean meridional flow velocity were more like 10 m s$^{-1}$ as 
suggested by feature tracking \cite[e.g.][]{hatha10}, then our 
quantitative estimates for ${\cal T}$ in Figure \ref{fig:FG}$a$ 
and $A_\phi$ in Figures \ref{fig:gyan2d}$c$ and \ref{fig:gyan} 
would be a factor of two too large.  Furthermore, if the actual 
amplitude of $\left<v_\theta\right>$ were to drop to zero near 
the base of the NSSL as suggested by \cite{hatha11b}, then our 
extrapolated values for $W_\phi$, $W_g$, and $W_t$ listed in 
Table 1 would be overestimated. 

With these variations and uncertainties in mind, we believe that
our quantitative estimate for $W_\phi$ given in \S\ref{sec:anen} 
is reasonable.  In particular, since the meridional flow speed 
$\left<v_\theta\right>$ is not likely to exceed 20 m s$^{-1}$ 
through much of the NSSL ($r > 0.95 R$), then the extrapolated 
value of $W_\phi$ is unlikely to exceed four percent of the solar 
luminosity, $L$.  Moreover, since the average value of 
$\left<v_\theta\right>$ throughout the NSSL ($r > 0.95 R$) 
is not likely to be much less than 10 m s$^{-1}$, then $W_\phi$ 
is unlikely to be less than 1-2 percent of $L$.

Our estimate for ${\cal G}$ is based solely on global 
rotation inversions from 1996.  Still, we can use the local
inversions for the zonal flow shown in Figure 
\ref{fig:omega_helio} to estimate the temporal variation.
This yields a standard deviation of less than 15\% for 
latitudes less than 40$^\circ$.  The maximum standard 
deviation at the highest latitudes measured, 52.5$^\circ$
is 34\%.  This may be regarded as an upper limit to the
uncertainty in the quantitative estimates for 
${\cal G}$, $A_r$, $A_\theta$, $W_r$, and $W_\theta$ in 
sections \ref{sec:fghelio} and \ref{sec:anen}.  We do 
not consider the time variation of global rotation inversions 
here but we expect they would imply smaller variations 
of ${\cal G}$ by virtue of their greater accuracy at
higher latitudes and lower radii.

\section{The Nature of Turbulent Transport}\label{sec:transport}

\subsection{Turbulent Diffusion and ${\cal L}$ Mixing}\label{sec:dm}

Perhaps the simplest paradigm one can envision to potentially account
for the existence of the NSSL is that of turbulent diffusion.  In
particular, one might suppose that as the length and time scales of
the convection decrease drastically approaching the photosphere from
below, the effects of rotation and spherical geometry become
unimportant, so the convection becomes more homogeneous and isotropic,
at least in the horizontal dimensions.  Furthermore, given the small
scale of the convective motions relative to the scale of the
differential rotation and meridional circulation, one might expect
them to suppress shear.  The decrease in $\Omega$ near the surface may
then be attributed to the vanishing of the Coriolis-induced velocity
correlations that sustain the differential rotation deeper in the
convection zone \citep[in mean-field parlance, this would be the
non-diffusive component of the Reynolds stress tensor, known as the
$\Lambda$-effect; e.g.][]{rudig89}.  The NSSL would then be maintained
by the outward diffusion of angular momentum from the deep convection
zone together with the advection of angular momentum by the meridional
flow.  Is this paradigm consistent with observations?

An alternative paradigm introduced briefly in \S\ref{sec:intro}, is that
photospheric convection tends to conserve angular momentum locally.
This has been found repeatedly in numerical simulations of rotating 
convection in parameter regimes with weak rotational influence 
\cite[high Rossby number, see:][]{gilma77,gilma79b,hatha82,deros02,aurno07,augus11}.
In the absence of external forcing or boundary influences, turbulent mixing 
tends to establish a rotation profile such that ${\cal L}$ is uniform 
throughout the domain in question ($\Omega \propto \lambda^{-2}$).  This
is particularly evident in the recent simulations by \cite{augus11} 
that employ open boundary conditions in the vertical, permitting
convective motions to pass through the boundaries.

In this section we consider these two paradigms for turbulent transport
and investigate whether they are consistent with the helioseismic inversions
discussed in \S\ref{sec:helio}.  We will focus primarily on the angular momentum 
transport, represented by ${\cal F}$, and its implications with regard to the 
$\Omega$ profile.  However, we emphasize again that the $\Omega$ profiles
considered here can only be maintained with the help of corresponding 
meridional forces ${\cal G}$ or ${\cal B}$; these meridional forces 
are {\em required} to maintain rotational shear $\pd \Omega / \pd z$ 
{\em regardless} of the nature of ${\cal F}$.

For both paradigms, we express the net axial torque in terms of the divergence 
of a turbulent angular momentum flux: ${\cal F} = - \dv {\bf F}$.  This is
consistent with the hyperbolic nature of the compressible MHD equations
and reliably captures the various components of ${\cal F}$, 
including the Reynolds stress, Lorentz force, and viscous diffusion
(Appx.\ A).
 
In the case of turbulent diffusion, the form of the angular momentum
flux ${\bf F}$ is analogous to that for molecular diffusion but the 
effective turbulent viscosity $\nu_t$ is many orders of magnitude 
larger than the molecular value.  This yields
\begin{equation}\label{eq:Fvd}
{\cal F} = - \dv {\bf F}_t = \dv \left(\rho \nu_t \lambda^2 \del \Omega\right) ~~~.
\end{equation}
Note that the direction of the angular momentum flux is down the gradient of 
$\Omega$, ${\cal F}_t \propto - \del \Omega$, tending to suppress rotational 
shear.   By contrast, in our second paradigm, turbulence tends to mix angular 
momentum so we expect the flux to be down the gradient of ${\cal L}$:
\begin{equation}\label{eq:Fmx}
{\cal F} = - \dv {\bf F}_a = \dv \left(\rho \nu_a \del {\cal L}\right)  ~~~.
\end{equation}
where ${\bf F}_a \propto - \del {\cal L}$.  Given the ${\cal L}$ profile
in Figure 1$b$, equation (\ref{eq:Fmx}) produces an angular momentum
flux that is radially inward and poleward, as in high-Rossby number
convection simulations \citep[e.g.][]{deros02}.

We emphasize that the motivation for equation (\ref{eq:Fmx}) is purely 
phenomenological; this form is suggested by simulations of turbulent 
convection at high Rossby number.  By contrast, equation (\ref{eq:Fvd}) 
can be derived rigorously through the techniques of mean-field theory, 
assuming scale separation and that small-scale motions are homogeneous 
and isotropic.

We note also that the paper that introduced the concept of the near-surface shear
layer, Foukal \& Jokipii (1975; hereafter FJ75), touched on both paradigms.  
They represented the turbulent
angular momentum flux as a viscous diffusion as in equation (\ref{eq:Fvd}) with
a constant dynamic viscosity $\rho \nu_t$.  Substituting this into equation (\ref{eq:gp})
and assuming a cylindrical rotation profile $\Omega = \Omega(\lambda)$ yields
\begin{equation}\label{eq:FJ}
\left<\rho v_\lambda\right> \frac{d {\cal L}}{d\lambda} = - \dv {\bf F}_t = \frac{\rho \nu_t}{\lambda}
\frac{d}{d \lambda} \left(\lambda^3 \frac{d \Omega}{d\lambda}\right)
~~~.
\end{equation}
It is straightforward to show that this equation is the same as equation (1)
in FJ75.  They then consider the advection-dominated limit
$\lambda v_\lambda >> \nu_t$ and argue that the flow will conserve
angular momentum, yielding a profile ${\cal L}$ = constant.

However, it is important to note that FJ75 attributed the conservation of angular
momentum to the {\em convection}, not to the mean flow.  If ${\cal L}$
represents the axisymmetric zonal flow, then only the axisymmetric meridional 
flow $\left<v_\lambda\right>$ (assuming 
$\left<\rho v_\lambda\right> \approx \left<\rho\right>\left<v_\lambda\right>$)
will contribute to the advection term on the left-hand-side (lhs) of 
equation (\ref{eq:FJ}).

Thus, there are two ways to interpret the FJ75 result within the context of
the two paradigms considered here.  The first interpretation is that the 
turbulent angular momentum flux is diffusive in nature [eq.\ (\ref{eq:Fvd})]
and the homogenization of angular momentum ${\cal L} \approx$ constant
is achieved by means of the meridional flow.  The second interpretation is
that the $\left<v_\lambda\right>$ and ${\cal L} = \lambda \left<v_\phi\right>$ terms on 
the lhs of equation (\ref{eq:FJ}) represent not mean flows but rather 
convection; that is, replace the mean velocities with fluctuating velocities.
When averaged over longitude and time, the net turbulent stress will then
tend to homogenize ${\cal L}$, as expressed in (\ref{eq:Fmx}).  

Yet, both interpretations are incomplete in the sense that they
do not take into account the observed poleward merdional flow which 
clearly crosses ${\cal L}$ contours as discussed in \S\ref{sec:helio}.
The meridional flow does not conserve angular momentum on its own
($\left<\rho \vv_m\right> \bdot \del {\cal L} \neq 0$), nor does
the convection (${\cal F} \neq 0$).  Both must contribute
to the subtle dynamical balances in the NSSL.

\subsection{Homogeneous Solutions and Stability}\label{sec:homo}

In \S\ref{sec:dm} we suggested two potential idealized forms for the
turbulent angular momentum flux, expressed in equations (\ref{eq:Fvd})
and (\ref{eq:Fmx}).  We now ask whether these paradigms are supported
by helioseismic inversions.  In this section we neglect the meridional
flow and consider only the homogeneous solutions discussed in 
\S\ref{sec:clarify}, equation (\ref{eq:homo}).  Namely, we calculate
the $\Omega$ profile that would result in no net axial torque,
${\cal F} = 0$ and we ask whether this profile resembles the 
$\Omega$ profile deduced from helioseismology.  Such solutions
satisfy our zonal momentum equation (\ref{eq:gp}) 
for $\left<\rho \vv_m\right> = 0$, as they must if they are to
describe a steady state.  We consider the influence of a 
meridional flow in \S\ref{sec:mflow}.

We begin with the case of turbulent diffusion and we assume for 
simplicity that the density-weighted (dynamic) diffusion coefficent
$\rho \nu_t$ is constant.  Anisotropic and inhomogeneous diffusion
coefficients will be considered in \S\ref{sec:anisotropy}.
From equation (\ref{eq:Fvd}), the condition
${\cal F} = 0$ then requires
\begin{equation}\label{eq:vdeq}
\dv \left(\lambda^2 \del \Omega\right) = 0 
\mbox{\hspace{0.2in} (turbulent diffusion).}
\end{equation}
Similarly, if we assume $\rho \nu_a$ is constant, equation
(\ref{eq:Fmx}) yields
\begin{equation}\label{eq:mxeq}
\nabla^2 {\cal L} = 0
\mbox{\hspace{0.2in} (${\cal L}$ mixing).}
\end{equation}

In order to obtain solutions to equations (\ref{eq:vdeq}) and
(\ref{eq:mxeq}) we must specify boundary conditions.  Thus, to
proceed, we must keep in mind the context in which these equations
are proposed to be valid.  They are intended to represent angular
momentum transport in the NSSL alone; angular momentum in the
deep convection zone (CZ) must be very different in order to sustain
the solar differential rotation.  Thus, to take into account the
coupling between the NSSL and the deep CZ, we specify boundary 
conditions at the base of the NSSL.

It is often stated that the $\Omega$ contours at mid latitudes
in the bulk of the CZ are nearly radial.  Close scrutiny of 
Figure \ref{fig:DR}$a$ reveals that $\Omega$ contours are not strictly 
radial; rather, they are tilted slightly toward the rotation axis 
so $\pd \Omega /\pd r > 0$ in the bulk of the convection zone.
However, in the NSSL, $\pd \Omega / \pd r < 0$.  Thus, we
define the base of the NSSL, $r_s$, as the radius at which
the spherically-averaged radial $\Omega$ gradient passes 
through zero.  This yields $r_s = 0.946$.

\begin{figure*}
\centerline{\epsfig{file=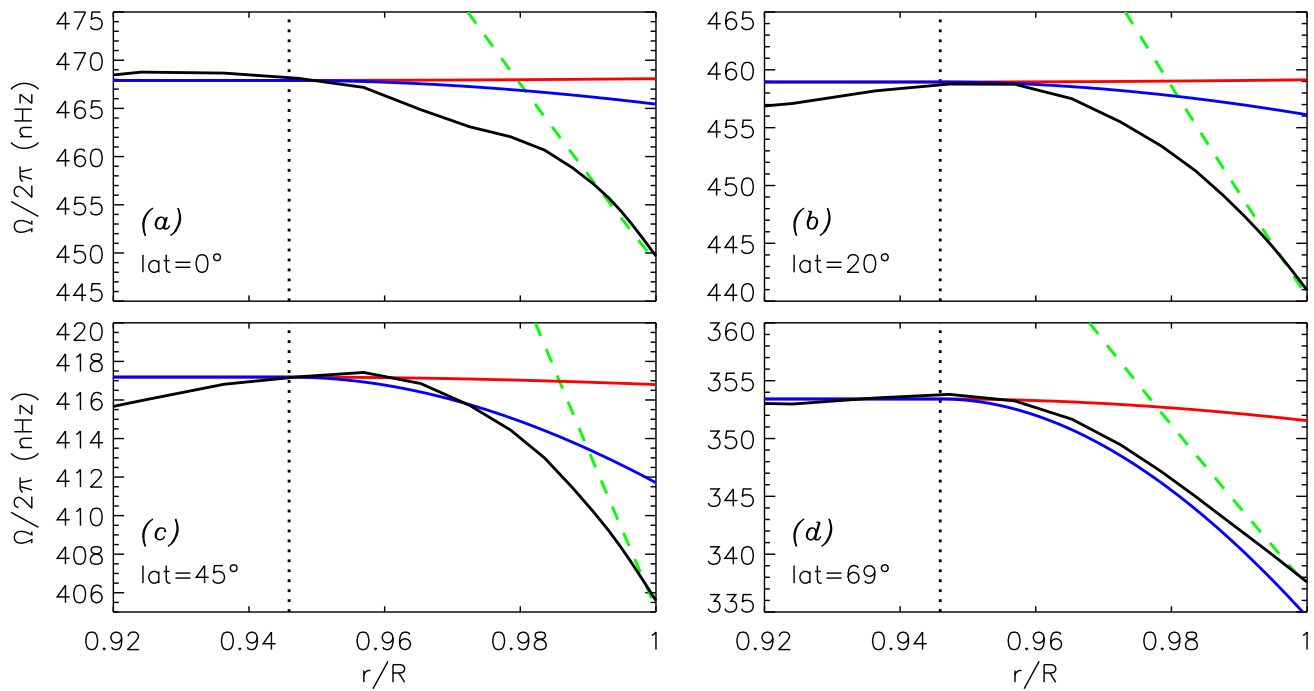,width=\linewidth}}
\caption{Rotation rate profiles are shown versus radius, spanning
the upper convection zone ($0.92 R \leq r \leq R$) for
latitudes of ($a$) 0$^\circ$, ($b$) 20$^\circ$, ($c$) 45$^\circ$, 
and ($d$) 69$^\circ$.  Dot-dashed lines (red in online color 
version of Figure) and
dashed lines (blue in online version) represent viscous diffusion
and ${\cal L}$ mixing respectively, obtained by solving equations
equations (\ref{eq:vdeq}) and (\ref{eq:mxeq}) subject to the
boundary conditions discussed in the text.  Solid lines (black 
in online version) represent the helioseismic 
inversions shown in Fig.\ \ref{fig:DR} and vertical dotted 
lines indicate the matching layer, $r=r_s$.  The sloped dotted
line (green dashed line in online version) represents an
angular momentum profile that is independent of depth: 
${\cal L} = {\cal L}_R(\theta)$ where 
${\cal L}_R(\theta)$ is the surface value (at $r = R$).
\label{fig:dp}}
\end{figure*}

Thus, we solve equations (\ref{eq:vdeq}) and (\ref{eq:mxeq}) in the
region $r_s \leq r \leq R$, subject to the boundary conditions 
$\pd \Omega / \pd r = 0$ and $\Omega = \Omega_s(\theta)$ at 
$r=r_s$.  Here $\Omega_s(\theta)$ is an analytic fit as in 
equation (\ref{eq:omega}) to the helioseismic inversions shown 
in Figure \ref{fig:DR}, interpolated to $r = r_s$.  For further details
on the boundary conditions and for the analytic solution of 
equations (\ref{eq:vdeq}) and (\ref{eq:mxeq}) see Appendix D.
The results are plotted in Figure \ref{fig:dp}.

It is immediately apparent in Figure \ref{fig:dp} that the actual
$\Omega$ profile in the Solar NSSL inferred from helioseismology
is steeper than suggested by either of the simple paradigms 
considered here.  Not surprisingly, viscous diffusion tends
to suppress shear so the equilibrium $\Omega$ profile in 
this case is nearly independent of radius.  However, the 
latitudinal differential rotation is still prominent, roughly
the same as at $r = r_s$ as a consequence of the thinness
of the layer (coupled with the boundary conditions).  Furthermore,
a weak negative radial shear ($\pd \Omega / \pd r$) is 
established at high latitudes.  This is a geometric effect 
associated with the inefficiency of viscous transport near the
rotation axis ($\lambda = 0$).

The profile corresponding to ${\cal L}$ mixing [eq.\ (\ref{eq:mxeq})]
is steeper than the diffusive profile, again as expected, since
the direction of the flux ($\propto - \del {\cal L}$) has a component
that is radially inward.  However, note that, within the context of the
NSSL, {\em the concept of turbulent convection mixing angular momentum 
is not the same as turbulent convection establishing ${\cal L} = $ constant}.
The thin-shell geometry and the coupling of the NSSL to the rotation
profile in the deep CZ (modeled here by means of our boundary conditions)
preclude a complete homogenization of ${\cal L}$ even if the local 
convective transport were to exhibit that tendency.  

At high latitudes, the $\Omega$ profile implied by ${\cal L}$ mixing
is steeper and is more comparable to the helioseismic inversions.
Both are less steep than the profile that would arise from 
a complete homogenization of ${\cal L}$.  However, the helioseismic
profile does approach homogenization in radius within about 7 Mm
of the photosphere.  In other words, the upper bound on the magnitude 
of $\pd \Omega/\pd r$ near the photosphere appears to be set by the 
Rayleigh stability criterion
 \citep{tasso78}
\begin{equation}\label{eq:ray}
\frac{\pd {\cal L}}{\pd \lambda} > 0 
\mbox{\hspace{.5in}($\del S = 0$)}  ~~~.
\end{equation}

Note that equation (\ref{eq:ray}) is only strictly valid under the assumption
that the stratification in the convection zone is approximately
adiabatic ($\del S = 0$).  This is of course an oversimplication
since the stratification in the NSSL is thought to be substantially
superadiabatic.  A more rigorous analysis indicates that the 
Rayleigh criterion (\ref{eq:ray}) is part of a more general 
formulation of the Schwarzchild criterion for convective stability, 
which is clearly violated in a convection zone, essentially 
by definition.  An independent stability criterion
requires $\cos\theta (\del S \cross \del {\cal L}) > 0$.  
However, if $S$ isosurfaces are predominantly horizontal, 
then this provides no constraints on the radial 
shear in the NSSL\footnote{In fact, it simply reflects
convective instability since 
$\cos\theta (\pd S/\pd r) (\pd {\cal L}/\pd \theta) < 0$}, 
$\pd \Omega / \pd r$.  
Both criteria arise from the Solberg-H{\o}iland stability 
analysis as discussed, for example
by \cite{tasso78}.  By considering the implications of 
equation (\ref{eq:ray}) here, we are essentially
separating out inertial effects from thermal effects, which
may operate on different time scales.  The potential relevance
of this separation to the NSSL is supported by the remarkable 
correspondence between the maximum radial gradient of the 
helioseismic $\Omega$ profile and the slope implied by 
equation (\ref{eq:ray}), as illustrated in Figure \ref{fig:dp}.

Thus, with these caveats, equation (\ref{eq:ray}) implies that 
the radial $\Omega$ gradient near the photosphere is what it is 
because anything steeper would be unstable.  Although this is 
indeed a compelling argument, it is well known that in the presence 
of a weak magnetic field the Rayleigh stability criterion is replaced 
by the stability criterion associated with the magneto-rotational
instability (MRI).  Thus, equation (\ref{eq:ray}) is replace by 
the condition that $\pd \Omega / \pd \lambda > 0$ 
\citep{balbu95}.  This condition is clearly violated in the NSSL.

Why might the rotation profile be limited by the hydrodynamic
Rayleigh criterion (\ref{eq:ray}) yet violate the MRI stability
criterion?  There are two potential answers to this question.  The first is
that the MRI stability analysis may not be applicable in the NSSL.
According to \cite{balbu95}, the assumptions that underlie the
MRI criteria are valid for field strengths in the range 
$4\pi\rho\chi\Omega << B^2 << 4\pi\rho R^2 \Omega^2$, where $\chi$
is the transport coefficient (in units of cm$^2$ s$^{-1}$) corresponding
to the dominant diffusive process.  Dissipation in the NSSL is likely
dominated by radiation and ohmic diffusion.  The detailed physics is 
complicated and very sensitive to depth but nevertheless, rough estimates 
for the corresponding diffusion coefficients give $\kappa \sim $10$^4$--$10^5$
cm$^2$ s$^{-1}$ and $\eta \sim 10^6$--$10^7$ cm$^2$ s$^{-1}$.  The 
$\kappa$ value is derived from solar structure Model S \citep{chris96} and the $\eta$ 
value follows from the Spitzer expression for a fully ionized Hydrogen 
plasma $\eta \sim 8 \times 10^{13} T^{-3/2}$ \citep{spitz62}.  Using
$\rho \sim 10^{-3}$ g cm$^{-1}$, we estimate that the MRI analysis
of \cite{balbu95} is valid for field strengths ranging from less than 
1 G to more than 10$^4$G.  The typical field strength in the NSSL 
is likely to lie within these bounds.

The second possibility is that the hydrodynamic Rayleigh instability 
is somehow more robust or more efficient than the MRI.  Thus, the 
NSSL may indeed be unstable to MRI but the time scale of the 
instability (of order the rotation period $\sim $ 28 days) 
is longer than the  time scale over which the shear is 
established by turbulent stresses (of order the convective turnover 
time $\sim $ 5 min -- 1 day).  Although this is plausible, 
the same argument would also in principle apply to the 
Rayliegh criterion so it is unclear why equation 
(\ref{eq:ray}) must be satisfied while the MRI criterion 
is not.  The depth over which the marginal slope is achieved,
roughly within $\sim $ 7 Mm of the photosphere, suggests that
granulation may play a role.  In any case, this is an interesting 
issue that should be explored further with the help of MHD convection 
simulations.

\subsection{Balancing the Meridional Flow}\label{sec:mflow}

The analysis presented in \S\ref{sec:homo} assumes that the net
torque ${\cal F}$ is zero, which is clearly not the case;
as demonstrated in \S\ref{sec:helio}, observations imply 
that ${\cal F}$ is negative.  Thus, we must take this into 
account if we are to properly assess whether our two simple 
paradigms for angular momentum transport can adequately 
account for the observed rotation and meridional flow profiles.

Stated another way, observations imply that the meridional flow is supplying
angular momentum to the NSSL, tending to speed up the local rotation
rate, while the turbulent transport must be removing angular momentum
in order to maintain a stationary state.  In order to assess whether
turbulent diffusion or ${\cal L}$ mixing can provide the requisite
transport, we must compute ${\cal F}$ from equations (\ref{eq:Fvd}) 
and (\ref{eq:Fmx}) based on the rotation profile inferred from 
helioseismology $\Omega_*(r,\theta)$ and ask whether it matches
the inferred net axial torque shown in Figure \ref{fig:FG}$a$.

\begin{figure}
\centerline{\epsfig{file=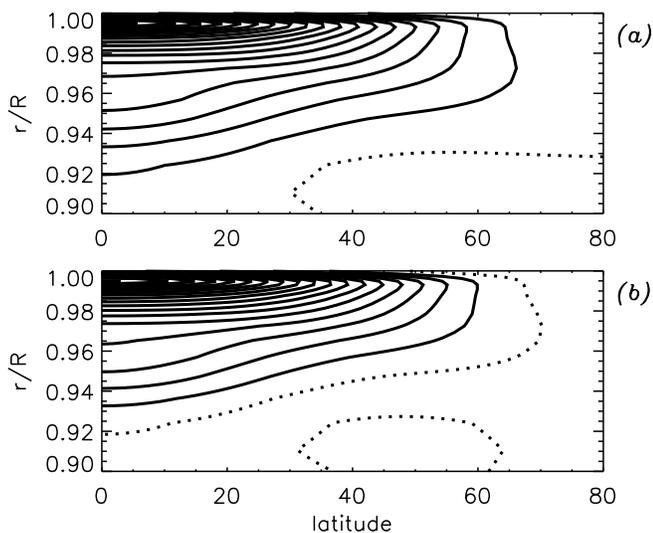,width=3.5in}}
\caption{Differential operators applied to the (smoothed)
solar $\Omega$ profile inferred from helioseismology
(Fig.\ \ref{fig:DR}), corresponding to ($a$) viscous 
diffusion $\dv (\lambda^2 \del \Omega_*)$ and 
($b$) ${\cal L}$ mixing, $\nabla^2 (\lambda^2 \Omega_*)$.  
Solid lines denote negative values, dotted lines denote zero and
positive values.  The scaling is arbitrary since the correspoding
torques depend on the unknown values of $\nu_t$ and 
$\nu_a$.\label{fig:lap}}
\end{figure}

Of course, we cannot compute ${\cal F}$ explicitly because we do 
not know {\em a priori} the transport coefficients $\nu_t$ and
$\nu_a$.  However, if we assume as in \S\ref{sec:homo} that
$\rho \nu_t$ and $\rho \nu_a$ are constant, then 
the corresponding torques, ${\cal F}_t$ and ${\cal F}_a$ 
will be proportional to the appropriate differential operators 
applied to the solar rotation profile, namely
\begin{equation}\label{eq:FtFa}
{\cal F}_t \propto \del \left(\lambda^2 \del \Omega_*\right)
\mbox{\hspace{.1in} and \hspace{.1in}}
{\cal F}_a \propto \nabla^2 \left(\lambda^2 \Omega_*\right) ~~~.
\end{equation}
The right-hand-side of each of these equations is plotted in
Figure \ref{fig:lap}.

The immediate impression from Figure \ref{fig:lap} is that the two
results look nearly identical.  This is a consequence of the
steepness of the radial $\Omega$ gradient and the thin-shell
geometry.  Thus, the dominant contribution to both operators 
is the second radial derivative:
\begin{equation}
\del \left(\lambda^2 \del \Omega_*\right) \approx \nabla^2 \left(\lambda^2 \Omega_*\right) 
\approx \lambda^2 \frac{\pd^2 \Omega_*}{\pd r^2}  ~~~.
\end{equation}
As shown in Figure \ref{fig:dp}, the solar $\Omega$ profile 
is nearly flat at the base of the NSSL ($\pd \Omega /\pd r \approx 0$)
and steepens with increasing radius, reaching its maximum slope
near the photosphere.  Thus, $\pd^2\Omega/\pd r^2 < 0$ and the
quantities shown in Figure \ref{fig:lap} are predominantly
negative in the NSSL ($r > 0.95R$).  

The negative sign of the quantities plotted in Figure \ref{fig:lap} 
bodes well for the viability of our two simple paradigms for angular 
momentum transport.  Both viscous diffusion and ${\cal L}$ mixing 
would tend to flatten out the negative curvature of the solar 
$\Omega$ profile ($\pd^2 \Omega/\pd r^2 < 0$).  This would tend
to decelerate the local rotation rate (${\cal F} < 0$) and
may thus serve to remove the angular momentum supplied to the
NSSL by the meridional flow.  The efficiency of the transport 
scales with the coefficients $\nu_t$ and $\nu_a$ and could
be calibrated to give the proper net torque, at least in
an integrated sense.

However, the profiles shown in Figure \ref{fig:lap} are clearly
different than the net axial torque profile inferred from helioseismic
measurements, shown in Figure \ref{fig:FG}$a$.  In particular, the
idealized profiles in Figure \ref{fig:lap} peak at the equator while
the helioseismic profile in Figure \ref{fig:FG}$a$ peaks at mid
latitudes, with a weak positive signal at the equator.  Thus, our
simple paradigms appear to be inconsistent with helioseismic
inversions.  However, it may plausibly be argued that the assumption of
homogeneous, isotropic coeffients $\rho \nu_t$ and $\rho \nu_a$ is
unrealistic and unnecessary.  In the next section we consider whether
this can redeem our simple paradigms.

Before proceeding, we note in passing that the 
${\cal L}$ mixing operator shown in Figure \ref{fig:lap}$b$
suggests a reversal in the sign of the torque ${\cal F}$
at high latitudes.  Since the local helioseismic inversions
only extend up to latitudes of about 50$^\circ$, we cannot
determine whether a similar sign change occurs in the 
Sun.  However, some measurements of the meridional flow
suggest that there may be persistent high-latitude 
counter-cells, which are particularly apparent at solar 
minimum when magnetic effects can be more easily separated 
out \cite[e.g.][see also Fig.\ \ref{fig:vth_helio}]{ulric10}.  
Under the justified assumption
that the angular momentum continues to decrease toward the 
poles ($\pd {\cal L}/\pd \theta < 0$ in the northern hemisphere),
equation (\ref{eq:gp}) implies that a high-latitude counter-cell
would indeed correspond to a change in sign of the net axial
torque ${\cal F}$ (such that ${\cal F} > 0$).

\subsection{Anisotropy and Inhomogeneity}\label{sec:anisotropy}

The idealized rotation profiles and net axial torques considered 
in \S\ref{sec:homo} and \S\ref{sec:mflow} are only valid for 
constant density-weighted transport coefficients 
$\rho \nu_t$ and $\rho \nu_a$.  If one allows for 
inhomogeneous, anisotropic diffusion tensors then can 
the two paradigms considered here provide a better fit
to helioseismic inversions?  

This is to some extent a tautology; one might expect that
one can construct a diffusion tensor to reproduce any
arbitrary $\Omega$ and ${\cal F}$ profiles so the results 
of this analysis would not be very enlightening.  However, 
we will demonstrate that turbulent diffusion can be ruled out 
as a viable paradigm even if it is anisotropic and inhomogeneous.  
Mixing angular momentum, on the other hand, may be 
more promising.

Note that, with regard to anisotropy, we are referring explicitly 
to the off-diagonal components of the viscous stress tensor 
$\nu_t$ or its ${\cal L}$-mixing analogue, $\nu_a$.  This should not
be confused with the off-diagonal components of the Reynolds
stress tensor, which are nonzero even for constant scalar
values of $\nu_t$ and $\nu_a$.  Indeed, the off-diagonal
components of the Reynolds stress tensor are essential
in order to account for the maintenance of mean flows.

Perhaps the most conspicuous fault with the profiles derived
in \S\ref{sec:homo} is that they admit an angular momentum
flux through the solar surface.  In the Sun, the sharp 
drop in density near the surface precludes any significant 
angular momentum flux through the photosphere on a time
scale comparable to the convection turnover time or the
rotation period.  Angular momentum is lost through the
solar wind but this loss rate is many orders of magnitude 
smaller than the rate at which angular momentum is 
continually circulated through the NSSL by convection 
and meridional flows.

Thus, we consider equation (\ref{eq:Fvd}) again but we now allow for 
a diffusion coefficient that depends on radius.  Furthermore,
we allow for anisotropic transport so we can regard $\nu_t$
as a tensor.  For the time being we will neglect the off-diagonal
elements of the tensor so we can express $\nu_t$ in terms
of vertical and horizontal component $\nu_{tv}$ and $\nu_{th}$.

For $\rho \nu_{tv}$ = constant we found in \S\ref{sec:homo} that
the resulting $\Omega$ profile was too shallow to account for 
the helioseismic $\Omega$ profile.  Can diffusion give rise to
a steeper profile?  The answer is yes, but only if the vertical
diffusion $\rho \nu_{tv}$ increases with radius 
($\pd (\rho \nu_{tv})/\pd r > 0$).  This would provide an 
outward angular momentum flux that becomes larger closer 
to the solar surface, providing the requisite flux divergence
to decelerate the NSSL.  However, as pointed out in the previous
paragraph, this cannot be sustained all the way to the photosphere
where ${\bf F}$ and thus $\rho \nu_{vt}$ must drop to zero.

To demonstrate that even an anisotropic, inhomogeneous diffusion
cannot account for the NSSL consider a closed volume $V$ bounded 
from below by the surface $r = r_b$ and from above by the 
photosphere $R$.  Latitudinal boundaries are at the equator and at 
an arbitrary colatitude $\theta_0$, located in the northern hemisphere 
close enough to the equator to be accessible to helioseismic inversions, 
say $\pi/6 < \theta_0 < \pi/2$ (corresponding to a latitude between 
0 and 60$^\circ$).  We now integrate the net axial torque ${\cal F}$ over
the volume, allowing for anisotropic transport and assuming 
${\bf F} = 0$ at $r=R$.  The symmetry of the $\Omega$ profile implies
no angular momentum flux through the equatorial plane 
($\pd \Omega / \pd \theta = 0$) so the integrated torque is given by
\begin{eqnarray}\label{eq:intor}
& & \int_V {\cal F} dV = - \int_{\cal S} {\bf F} \bdot d\surf \\
& = & - 2 \pi \left[r_b^4 \int_{\theta_0}^{\pi/2} \rho \nu_v \frac{\pd \Omega}{\pd r}
\sin^3\theta d\theta + \int_{r_b}^R \rho \nu_h \frac{\pd \Omega}{\pd \theta}
r^2 dr \right] \nonumber
\end{eqnarray}
where ${\cal S}$ is the bounding surface, and the first and second
terms on the right-hand-side are to be evaluated at $r=r_b$ and
$\theta = \theta_0$ respectively.  

As discussed in \S\ref{sec:helio} and \S\ref{sec:mflow}, the
integral in equation (\ref{eq:intor}) must be negative.
However, if we choose $r_b$ to lie within the NSSL, then 
$\pd \Omega / \pd r < 0$ and the first term is positive.
In other words, angular momentum flux into the NSSL through the lower
surface $r = r_b$ must be shunted poleward in order to be consistent with a
negative (or zero) net torque $\int_V {\cal F} dV$.  Although the latitudinal
flux is indeed poleward ($\pd \Omega / \pd \theta < 0$ at $\theta =
\theta_b$), it cannot be efficient enough to maintain a negative $\pd
\Omega/\pd r$ throughout the NSSL.  As $r_b$ approaches $R$ from
below, the second term in brackets can be approximated by $\rho \nu_h
R^2 (R-r_b) (\pd \Omega/\pd \theta)$.  In order to transport the
requisite flux, the viscosity anisotropy $\nu_h/\nu_v$ would have to
increase radially without bound, becoming infinite at the photosphere.

This argument can be readily generalized to rule out any arbitrary
outward flux ${\bf F} \bdot \uvr > 0$.  The implication, then,
is that {\em the angular momentum flux in the NSSL must be radially
inward}.  One can in principle salvage the diffusive paradigm
if one invokes negative diffusion or the off-diagonal elements 
of the turbulent viscosity tensor, namely an inward angular 
momentum flux that is proportional to the latitudinal shear.  
However, such prescriptions seem rather contrived.

Thus, we can confidently say that the turbulent transport in the
NSSL cannot be adequately modeled as a turbulent diffusion.
How does our other paradigm fare, namely that of mixing angular 
momentum?  

We can see immediately that this paradigm is more plausible because
the angular momentum transport is indeed inward 
${\bf F} \bdot \uvr < 0$.  A steeper $\pd \Omega / \pd r$
profile could then be achieved by means of a diffusion coefficient
$\rho \nu_a$ that decreases with radius ($\rho \nu_a$), approaching
zero at the photosphere, as required by the condition ${\bf F} = 0$
at $r = R$.  Thus, the ${\cal L}$-mixing paradigm is consistent
with both an inward $\Omega$ gradient ($\pd \Omega/\pd r < 0$) 
and no net angular momentum flux through the photosphere.

However, in order to be consistent with the form of the net axial
torque inferred from helioseismology, the density-weighted mixing
coefficient $\mu = \rho \nu_a$ would have to have a very particular
form.  As shown in Figure \ref{fig:FG}$a$, the net axial torque
is nearly zero at the equator whereas $\nabla^2 {\cal L}$ has
a prominent peak at the equator (Fig.\ \ref{fig:lap}$b$).
In order to see how this may be remedied, assume that $\mu$
is isotropic but varies with radius with a scale height much
less than the solar radius $R$.  Furthermore, assume
as in equation (\ref{eq:cylindrical}) that the ${\cal L}$ 
profile is nearly cylindrical.  Then equation (\ref{eq:Fmx})
yields
\begin{equation}\label{eq:inhomo}
{\cal F} \approx \mu \nabla^2 {\cal L} + 2 \sin\theta \frac{d\mu}{d r} \frac{d {\cal L}}{d\lambda}
+ {\cal L} \frac{d^2 \mu}{dr^2} ~~~.
\end{equation}
If $\mu = \rho \nu_a$ decreases with increasing radius as 
suggested in the previous paragraph, then inserting the 
helioseismic value for ${\cal L} = \lambda^2 \Omega_*$ 
implies that the first two terms are negative at the
equator.  In other words, the sharp decrease of $\mu$
exacerbates the prominent negative net torque ${\cal F}$ 
seen in Figure \ref{fig:lap}$b$.  One way out of this
dilemma is if the slope flattens out near the photosphere
so $d^2\mu/dr^2 > 0$.  Then the last term in 
equation (\ref{eq:inhomo}) is positive and correspondence 
with Figure \ref{fig:FG}$a$ becomes possible.

In summary, helioseismology sets fairly strict constraints on the
nature of the turbulent angular momentum flux ${\bf F}$ in the NSSL, 
including: (1) ${\bf F}$ must be radially inward (${\bf F} \bdot \uvr < 0$),
(2) there must be no flux through the photosphere 
(${\bf F} \bdot \uvr = 0$ at $r = R$), (3) ${\bf F}$ must 
diverge at mid-latitudes (${\cal F} < 0$) in order to balance the 
advection of angular momentum by the meridional flow, and
(4) turbulent transport at low latitudes must redistribute
angular momentum in such a way as to support a steep
radial $\Omega$ gradient while minimizing the net 
axial torque (${\cal F} = \dv {\bf F} = 0$).
The turbulent diffusion paradigm, equation (\ref{eq:Fvd}), 
does not meet these constraints and is therefore not
a valid model of turbulent transport in the NSSL.  
The alternative ansatz of ${\cal L}$ mixing, as expressed 
in equation (\ref{eq:Fmx}), is at least feasible, but it 
would require a fine-tuned, inhomogeneous and/or anisotropic
mixing coefficient $\mu = \rho \nu_a$.  For example, it
could in principle be achieved with a vertical mixing coefficient 
that decreases with radius but flattens out near the photosphere
($\pd \mu/\pd r < 0$, $\pd^2 \mu/\pd r^2 > 0$). A power law
dependence $\mu \propto r^n$ with $n < 0$ may satisfiy such
a requirement.  However, it is more likely that the turbulent
Reynolds stress in the NSSL is more complex than either of
these crude, local models.

\subsection{Meridional Momentum Transport}

Despite the conclusion of section \S\ref{sec:anisotropy} regarding the
non-diffusive nature of ${\cal F}$, it is reasonable to expect that
${\cal G}$ may operate essentially as a turbulent diffusion.  In order
to appreciate why this may be the case, consider again the 
time-dependent thought experiment discussed in \S\ref{sec:thex}.  
Here we begin with a spherical volume $V$ in uniform rotation and 
a retrograde torque ${\cal F}$ is introduced in the surface layers 
($r > 0.95$).  This will establish a poleward flow $\left<v_\theta\right>$
which will steadily increase in amplitude, striving to 
establish a cylindrical rotation profile as discussed in
\S\ref{sec:thex}.  Given the thin-shell
geometry of the NSSL, one may expect strong vertical
gradients $\pd v_\theta /\pd r$ to be established.  Furthermore, 
given the small scale of photospheric convection relative to the mean flow,
one may expect turbulent stresses to resist such shearing motions. 
This would imply that the resulting axial differential rotation 
profile $\pd \Omega / \pd z$ is determined by how efficiently small-scale
convective motions can mix latitudinal momentum, $\left<v_\theta\right>$, 
or equivalently, suppress zonal vorticity, $\left<\omega_\phi\right>$.

Thus, under this scenario we may expect that ${\cal G}$ will be 
diffusive (down-gradient) in nature, and furthermore, that the strain 
rate tensor will be dominated by vertical gradients in the poleward 
flow.  The turbulent transport would then be given by
\begin{equation}\label{eq:tdiff}
{\cal G} \sim \nu_t \frac{\pd^2 \omega_\phi}{\pd r^2} \sim
\nu_t \frac{\pd^3 v_\theta}{\pd r^3} ~~~,
\end{equation}
where $\nu_t$ is again the turbulent (kinematic) viscosity.  Note that this
expression assumes that the scale of variation of the dynamic viscosity
$\vert d \ln (\rho \nu_t)/d r \vert^{-1}$ is larger than that of the shear
$\vert d \ln \vert v_\theta\vert / dr \vert^{-1}$.

Equation (\ref{eq:tdiff}) can in principle provide the momentum 
transport required to balance the Coriolis force associated
with the axial shear $\pd \Omega/\pd z$.  However, this is 
contingent on the amplitude of $v_\theta$ being strongly peaked
near the photosphere, such that $\pd^3 \vert v_\theta \vert/\pd r^3 < 0$.
This does not appear to be supported by local helioseismic 
inversions (\S\ref{sec:helio}), but the sensitivity and resolution
of the inversions is likely not adequate enough to provide reliable
estimates for third-order derivatives of $v_\theta$
\cite[e.g.][]{becke07,hatha11b}.  An alternative is that the dynamic turbulent 
viscosity $\rho \nu_t$ decrease with radius while the velocity amplitude 
increase, consistent with no tangential stress at the photosphere.

In summary, we can say that if the turbulent transport ${\cal G}$ 
does indeed dominate over baroclinic forcing ${\cal B}$ as suggested
in \S\ref{sec:what} then it must resist the poleward meridional flow 
induced by ${\cal F}$ through the Coriolis force.  Thus, ${\cal G}$ 
must be negative in the northern hemisphere and positive in the 
southern hemisphere.  A turbulent diffusion or similar mixing 
process may be adequate, as would a more general formulation.  

\section{Summary and Conclusion}\label{sec:summary}

\subsection{Maintenance of the Solar NSSL}

We have demonstrated that the turbulent angular momentum 
transport in the solar Near-Surface Shear Layer (NSSL) is 
responsible for the persistent poleward meridional flow but it does
not uniquely determine the mid-latitude $\Omega$ profile.  
Rather, the axial rotation gradient $\pd \Omega/\pd z$ must be
maintained by turbulent stresses in the meridional plane.
More specifically, a retrograde zonal force (axial torque)
${\cal F} < 0$ establishes the NSSL and regulates the poleward
meridional flow while meridional forcing regulates the mid-latitude
$\Omega$ profile.  Furthermore, we argue that a transition in the 
meridional force balance from baroclinic to turbulent stresses 
[${\cal B}$ to ${\cal G}$ in equation (\ref{eq:mer})]
may determine the base of the NSSL.

More generally, we have demonstrated that there is close dynamical
relationship between the differential rotation and the meridional
circulation and that the structure of the NSSL as inferred from 
helioseismology relies on a delicate nonlinear, nonlocal interplay 
between the two.  The same physical mechanism (${\cal F} < 0$) that 
establishes negative radial shear $\pd \Omega/\pd r < 0$ also 
establishes poleward flow ($\left<v_\theta\right> < 0$ in the 
northern hemisphere).  As suggested by previous authors, we 
attribute this physical mechanism to Reynolds and possibly Maxwell 
stresses associated with the relatively small-scale convection 
(granulation to supergranulation) that permeates the NSSL.

Throughout our analysis, we have considered the inertia of the
mean flow explicitly, incorporating other physical processes
in the generalized zonal and meridional forcing terms 
${\cal F}$ and ${\cal G}$ which we refer to as turbulent stresses.
This is intended to clarify the essential physics of how the differential
rotation and meridional circulation are coupled in their response to
zonal and meridional forcing, independent of the detailed nature of
this forcing, which is unknown.  The reader may wish to regard ${\cal F}$ 
and ${\cal G}$ simply as the the convective Reynolds stress, since this
is likely to be their dominant component (Appendix A).  However, as
noted in the introduction, other forcing such as large-scale Lorentz
forces and viscous diffusion can induce mean flows in an analogous
way.

We have demonstrated that turbulent transport in the NSSL is
non-diffusive in nature and must be directed radially inward
(\S\ref{sec:transport}).  Inspired by numerical simulations of
turbulent convection at large Rossby numbers, we have considered an
alternative paradigm for turbulent transport based on the mixing of
specific angular momentum, ${\cal L}$.  We have shown that the
conservation of angular momentum alone cannot account for the
existence of the NSSL (\S\ref{sec:meridional}) but the form of the
rotation profile may be consistent with ${\cal L}$ mixing if the
mixing coefficient is inhomogeneous and/or anisotropic
(\S\ref{sec:transport}).  Furthermore, we have shown that ${\cal L}$
mixing is a more general concept than simply ${\cal L}$-homogenization
(constant ${\cal L}$) when one takes into account the coupling between
the NSSL and the deep convection zone.  Yet, the failure of simple
turbulent diffusion or ${\cal L}$-mixing prescriptions demonstrates
that the NSSL is not simply a passive response to deeper forcing; it
must be actively maintained by anisotropic and inhomogeneous turbulent
transport.

Estimates based on local and global helioseismology indicate
that it takes 2--4\% of the solar luminosity to maintain
the NSSL against the inertia of the mean flow
(\S\ref{sec:anen}).  Most of this work is associated with
transporting angular momentum out of the layer ($W_t$ in 
Table 1) in order to balance the convergence of angular 
momentum flux into the layer by meridional flow advection.
The estimated amplitudes of turbulent transport in the
latitude and longitude directions ($A_\theta$ and $A_\phi$)
are remarkably similar, given the very different way
in which these two quantities were obtained (the former
follows from ``uncurling'' global $\Omega$ inversions while
the latter involves estimates of zonal and meridional flows
from local helioseismology; see \S\ref{sec:anen}). The
vertical transport in the middle of the NSSL is about 
an order of magnitude less (at the base of the NSSL,
$r = r_s$, $A_\theta$ is zero by assumption and
our local helioseismic inversions provide no information
on $A_\phi$).  The sense of all these terms is such that 
turbulent transport is decelerating the rotation rate in 
the NSSL and opposing the meridional flow.

Estimates of the spin-down time scale indicate that it is similar
to the ventilation time scale of about 600 days (\S\ref{sec:anen}).
This is long compared to the turnover time scale of convective
motions, $\lesssim $ 1 day, implying that the net turbulent
angular momentum transport is rather inefficient.

Finally, we note that the upper limit to the slope of the
radial $\Omega$ gradient appears to be set by the 
Rayleigh criterion, equation (\ref{eq:ray}).  However, it
is unclear why this intrinsically hydrodynamic condition
should be satisfied in the NSSL while its MHD analogue, the 
stability criterion for the magnetorotational instability 
(MRI) $\pd \Omega / \pd \lambda > 0$, is clearly violated.

\subsection{Implications for Numerical Models}

Several authors have sought to investigate the dynamics of the 
NSSL through numerical simulations of solar convection in thin
spherical shells or spherical segments, placing the lower boundary
of the simulation domain within the upper convection zone, typically
above 0.9$R$  \cite[e.g.][]{gilma79b,hatha82,deros02,augus11}.  
Relative to global simulations spanning the entire convection zone,
these have the great advantage that smaller scales can be resolved 
so the turbulent transport can be more reliably captured.  
Although these simulations have provided insight into the dynamics
of the NSSL, none have accurately reproduced the angular velocity
profile throughout the NSSL.  As we have demonstrated here,
the dynamics of the NSSL involves a delicate balance between 
small-scale turbulent transport, large-scale mean flows, and 
coupling to the deep convection zone (CZ).  Even if thin-shell 
models properly capture the turbulent transport, they must also 
capture or otherwise mimic the coupling to the CZ if they are 
to achieve solar-like mean flows.

A straightforward and common strategy in thin-shell models is to
impose a solar-like latitudinal differential rotation on the lower
boundary \cite[e.g.][]{gilma79b,deros02}.  If the lower boundary is also
impermeable, as is often the case, then it is clear that the system
cannot sustain a poleward flow throughout the layer.
What implications might this have for the differential
rotation?  As is demonstrated in $\S\ref{sec:helio}$, the meridional
circulation supplies angular momentum to the NSSL while turbulent
stresses must remove this angular momentum.  This balance, together
with the meridional stresses ${\cal G}$ (or ${\cal B}$) determine the
mean flow profiles.  If the poleward flow is artificially suppressed
by an impenetrable boundary condition, then there must be alternative
source of angular momentum to balance turbulent transport.  Without
this source, the $\Omega$ profile will be adversely affected in
addition to the meridional flow profile.

The rotational coupling between the convection zone and the NSSL in
this class of numerical models is achieved by means of viscous
diffusion.  This is artificial in the sense that the viscosity used is
many orders of magnitude larger than that of the solar plasma.
However, can this effectively mimic the coupling that is expected to
occur in the Sun?  On the positive side, viscous coupling can indeed
serve as an angular momentum source, allowing the integrated turbulent
stresses $\int_V {\cal F} dV$ throughout the NSSL to be negative, as
in the Sun.  However, the latitudinal distribution of the viscous torques 
is likely to be very different than for gyroscopic pumping.  If 
$\pd \Omega/\pd r < 0$ at nearly all latitudes in
the solar NSSL as implied by helioseismic inversions (although these
are uncertain poleward of 70$^\circ$), then viscous coupling would
imply outward angular momentum transport everywhere.  By contrast, the
meridional flow would impart angular momentum at low latitudes and
remove it at high latitudes in such a way that the net convergence
into the layer is positive.  This implies a high-latitude angular momentum
flux that may be up the gradient of $\Omega$.  One would expect the 
resulting $\Omega$ profile to be very different than for viscous coupling.
Mean flow profiles would be similarly sensitive to impermable boundary 
conditions in latitude.

Thus, in order to properly model the NSSL, numerical models must either
be deep enough to capture the closed meridional circulation, including the
poleward flow at the surface and the return equatorward flow, or they
must impose boundary conditions that are conducive to establishing
solar-like mean flows.  These may include open boundaries on which the 
rotation profile $\Omega$ is specified in addition to an imposed net axial 
torque ${\cal F}$, chosen to induce a commensurate meridional flow through 
the boundary by means of gyroscopic pumping, as expressed by 
equation (\ref{eq:gp}).

Even if the simulation domain is in principle large enough to capture the
complete, closed meridional circulation, the delicate balance expressed in 
equation (\ref{eq:gp}) implies that the meridional flow is particularly
sensitive to artificial viscous dissipation.  If viscous angular momentum 
transport largely balances the angular momentum transport by the convective 
Reynolds stress as in many numerical models, then ${\cal F}$ will nearly 
vanish and the meridional flow may be very different than what occurs 
in the Sun \citep{miesc08,miesc11a}.  Regardless of the boundary conditions, 
realistic meridional flow profiles require minimal viscous dissipation.

\acknowledgements
We are very grateful to Rachel Howe for providing the global rotational
inversions shown in Figure \ref{fig:DR} and analysed in \S\ref{sec:helio}
and Mark Rast for providing data and insight with regard to photospheric 
irradiance variations.  We thank Michael Thompson and an anonymous referee
for comments on the manuscript 
and Kyle Auguston, Michael McIntyre, Matthias Rempel and Juri Toomre for many 
helpful discussions.  In particular, we thank Michael McIntyre for 
inspiring and educating us on the joys of gyroscopic pumping over the years
and for suggesting the term ``force curl'' for ${\cal G}$.  This work is 
supported by the NASA Heliophysics Theory Program, grant number NNX08AI57G 
as well as NASA grants NNH09AK14I (M.S.M.) and NNX08AJ08G, NNX08AQ28G 
and NNX09AB04G (B.W.H.).  
NCAR is sponsored by the National Science Foundation.


\appendix

\section{Appendix A: Explicit Expressions for ${\cal F}$ and ${\cal G}$}

In this Appendix we explicitly identify what is
included in the turbulent stress terms ${\cal F}$ and ${\cal G}$ that are introduced
in sections \ref{sec:concepts} and \ref{sec:zonvort} and that are used throughout 
the paper.

We begin with the equation that expresses the conservation of momentum in a compressible,
electrically conducting fluid under the magnetohydrodynamic (MHD) approximation
\begin{equation}\label{appa:mom}
\rho \frac{\pd \vv}{\pd t} + \rho \left(\vv \bdot \del\right) \vv = - \del P 
+ \rho \grav + \frac{1}{4\pi} \left(\curl \BB\right) \cross \BB + 
\dv \DD
\end{equation}
Traditional notation is used: $\vv$ is the bulk velocity of the fluid, $\rho$ is the density,
$P$ is the pressure $\grav = - g \uvr$ is the gravitational acceleration, $\BB$ is the magnetic
field, and $\DD$ is the viscous stress tensor, with elements
\begin{equation}\label{appa:visc}
{\cal D}_{i j} = - 2 \rho \nu \left[e_{i j} - \frac{1}{3}\left(\dv \vv\right)^2\right]
\end{equation}
where $\nu$ is the kinematic viscosity.  We use spherical polar coordinates ($r, \theta, \phi)$
and throughout the bulk of the paper, we consider an inertial reference frame.  This lets us
more gracefully incorporate the differential and uniform rotation components into a single,
non-uniform rotation profile, $\Omega$.  However, for the benefit of readers, in this Appendix
we wish to illustrate explicitly where the Coriolis force enters into this analysis.

Thus, we can convert equation (\ref{appa:mom}) into a rotating coordinate system 
($r$, $\theta$, $\phi^\prime$) by writing $\vv = \uu + \Omega_0 \lambda \uvp$ and 
$\phi_r = \phi + \Omega_0 t$.  Substituting 
these changes into (\ref{appa:mom}) yields
\begin{equation}\label{appa:rot}
\rho \frac{\pd \uu}{\pd t} + \rho \left(\uu \bdot \del \right) \uu = - \del P 
+ \rho \grav - 2 \rho \oom \cross \uu - \rho \oom \cross \left(\oom \cross \blam\right) +
\frac{1}{4\pi} \left(\del \cross \BB\right) \cross \BB + \del \bdot \DD_r
\end{equation}
where $\oom = \Omega_0 \uvz$, $\blam = \lambda \uvl$.  All derivatives 
are with respect to the rotating coordinate system so $\phi$ may be formally replaced by 
$\phi_r$ in the $\del$ operators.  However, this is not necessary since 
$\pd / \pd \phi_r = \pd / \pd \phi$ for fixed $r$, $\theta$, and $t$.  The stress tensor
$\DD_r$ is the same as $\DD$ with $\vv$ replaced by $\uu$ (and $\phi$ by $\phi_r$).

Multiplying the zonal component of (\ref{appa:rot}) by $\lambda$ and averaging over 
longitude ($\phi_r$) yields
\begin{equation}\label{appa:amom}
\frac{\pd }{\pd t}\left<\rho \lambda u_\phi\right> + \left<\rho \uu_m\right> \bdot \del {\cal L}
= {\cal F} \equiv 
- \dv \left[\left<\rho \lambda \uu^\prime u_\phi^\prime\right> - 
\left<\lambda \BB B_\phi\right> - \rho \nu \lambda^2 \del \Omega\right]
\end{equation}
where ${\cal L} = \lambda \left(\left<u_\phi\right> + \lambda \Omega_0\right)
= \lambda \left<v_\phi\right> = \lambda^2 \Omega$.
The right-hand-side is defined as the net axial torque ${\cal F}$.  Thus, it includes
the Reynolds stress (first term), the Lorentz force (second term), and the
viscous diffusion (third term).  The Lorentz force may be decomposed into
a contirbution from mean fields $\left<\BB\right>\left<B_\phi\right>$ and a Maxwell stress,
$\left<\BB^\prime B_\phi^\prime\right>$.  Meanwhile, the left-hand-side 
of (\ref{appa:amom}) includes the Coriolis force and the nonlinear advection (inertia)
associated with the mean flows, $\left(\left<\vv\right> \bdot \del \right) \left<\vv\right>$.

We emphasize that equation (\ref{appa:amom}) follows directly from equation (\ref{appa:rot}) 
with no additional assumptions [although we have used the mass continuity equation
$\pd \rho/\pd t = - \dv (\rho \uu)$ in the derivation].  Equation (\ref{appa:rot}) 
in turn follows directly from (\ref{appa:mom}).  The only assumption in any of this derivation 
is the MHD approximation that underlies equation (\ref{appa:mom}).  If we make the justified 
assumption that $\left<\rho u_\phi\right> \approx \left<\rho\right> \left<u_\phi\right>$ 
(as, for example, in the anelastic approximation), then the first term on the left-hand side 
of (\ref{appa:amom}) is just $\pd (\left<\rho\right> {\cal L})/\pd t$.  Furthermore, note
that $\uu_m = \vv_m$.  We thus obtain equation (\ref{eq:amom}).

Now consider the meridional components of (\ref{appa:rot}).  As is well known, the gravitational
and centrifugal terms can be expresed in terms of a gradient  
$\grav + \oom \cross (\oom \cross \blam) = \del (\Psi_g + \lambda^2 \Omega_0^2/2)$ where
$\Psi_g$ is the gravitational potential.  Furthermore, we may combine the advection and
Coriolis terms as follows
\begin{equation}
\left(\uu \bdot \del\right) \uu + 2 \oom \cross \uu = \vort \cross \uu + 
\del \left(\frac{u^2}{2}\right) ~~~.
\end{equation}
where $\vort = \curl \uu + 2 \oom = \curl \vv$ is the vorticity relative to the inertial
frame, also referred to as the absolute vorticity.  We can then divide (\ref{appa:rot}) by
$\rho$, average over longitude, and compute the zonal component of the curl to obtain
\begin{eqnarray}
\frac{\pd \left<\omega_\phi\right>}{\pd t} + \lambda \frac{\pd \Omega^2}{\pd z} & = &
{\cal B} + {\cal G} =  
\frac{\del \left<P\right> \cross \del \left<\rho\right>}{\left<\rho\right>^2} 
+ \left<\frac{\del P \cross \del \rho}{\rho^2} - 
\frac{\del \left<P\right> \cross \del \left<\rho\right>}{\left<\rho\right>^2}\right>
\nonumber \\
& + & \left\{\curl \left[\left(\curl \left<\uu_m\right>\right)\cross \left<\uu_m\right>
\right]\right\} \bdot \uvp
\nonumber \\
& + & \left\{\curl \left<\left(\curl \uu^\prime\right)\cross \uu^\prime
+ \frac{1}{4\pi\rho} \left(\del \cross \BB\right) \cross \BB + 
\rho^{-1} \del \bdot \DD_r \right>\right\} \bdot \uvp ~.
\label{appa:mer}
\end{eqnarray}
Again, this equation follows directly from equation (\ref{appa:rot}) with no further 
assumptions.  The second term on the left-hand-side, involving $\Omega^2$ includes 
the Coriolis force and the inertia associated with the differential rotation, 
$\left<u_\phi\right>$.  The term on the right-hand-side involving the mean meridional 
circulation, $\left<\uu_m\right>$ is estimated to be about two orders 
of magnitude smaller than this.  Since we wish to focus on the primary components 
that contribute to the force balance and the inertia of the mean flow, we include 
the meridional circulation term with the turbulent stress term ${\cal G}$.
Alternatively, since the kinetic energy density of the convection is at least
two orders of magnitude larger than that in the meridional circulation,
it is justified to neglect the meridional flow term in ${\cal G}$ altogether,
relative to the Reynolds stress. 

The first term on the right-hand-side of equation (\ref{appa:mer})
is the baroclinic term associated with the mean stratification, $\left<\rho\right>$ and
$\left<P\right>$, which we define as ${\cal B}$ in equation (\ref{eq:bc}).  The remaining
terms on the right-hand-side of equation (\ref{appa:mer}) define the turbulent stress
${\cal G}$.  Thus, ${\cal G}$ includes residual baroclinic forcing involving
fluctuating density and pressure components $\rho^\prime$ and $P^\prime$.  In fact, 
if we again make the approximation that $\rho \approx \left<\rho\right>$ to lowest 
order, then this residual baroclinic term is just proportional to 
$\left<\del \rho^\prime \cross \del P^\prime\right>$.
Other contributions to ${\cal G}$ include the nonlinear advection of the mean meridional 
flow (the term involving $\left<\uu_m\right>$, which is likely negligible as noted 
above), the Reynolds stress (the term involving $\uu^\prime$), the Lorentz force 
(the term involving $\BB$), and the viscous diffusion (the term involving $\DD$).

We use the term {\em turbulent stresses} loosely, since it is clear that ${\cal F}$ and
${\cal G}$ may in principle include contributions from the large-scale Lorentz force and 
the viscous diffusion, which need not be turbulent.  However, these are not likely to 
be important in the solar NSSL.  The small molecular viscosity of the solar plasma makes 
viscous diffusion negligible and the persistence of the NSSL throughout the solar cycle 
suggests that it is not maintained by the large-scale Lorentz force.  Thus, we expect 
that the dominant components of ${\cal F}$ and ${\cal G}$ are indeed the turbulent 
Reynolds and Maxwell stresses associated with small-scale convection.

\section{Appendix B: An Analytic Illustration of Gyroscopic Pumping}\label{appx:analytic}

Because it is essentially non-local and thus sensitive to the global geometry,
boundary conditions, and inhomogeneities, gyroscopic pumping is a subtle phenomenon
that generally requires numerical calculations to find equilibrium states.  Still,
analytic solutions can be found for idealized cases.  Here we present such an
analytic solution in the context of the solar NSSL.  For others, 
see \cite{hayne91}, \cite{garau08}, \cite{garau09}, and \cite{garau10}.

We emphasize again the point made in \S\ref{sec:concepts}; that the gyroscopic
pumping equation (\ref{eq:gp}) can induce a meridional flow with relatively
little impact on the rotational shear.  Here we illustrate this by independently 
specifying both a differential rotation profile $\Omega(\lambda)$ and a zonal 
force ${\cal F}$ and then proceeding to derive a solution for the meridional 
flow that links the two.

We restrict our attention to the northern hemisphere.  Solutions for the
southern hemisphere then follow by symmetry.  Consider a simple, cylindrical 
angular velocity profile given by
\begin{equation}\label{aomega}
\Omega = \Omega_0 + \Delta \Omega \frac{\lambda}{R} ~~~. 
\end{equation}
This, by construction, satisfies the steady state ($\pd \left<\omega_\phi\right>/\pd t = 0$) 
meridional force balance (eq.\ (\ref{eq:mer})) with ${\cal B} = {\cal G} = 0$.  In this
appendix we will find a steady solution $\Psi(\lambda,z)$ to the zonal force balance 
equation (\ref{eq:gp}) for the $\Omega$ profile in (\ref{aomega}) and a specified 
torque ${\cal F}$.  We will treat the differential rotation $\Delta \Omega$ as a free
parameter.  Note that this is not a unique solution in the sense that other
mean flow profiles $\Omega(\lambda)$, $\Psi(\lambda,z)$ can be realized with
the same ${\cal F}$.  The equilibrium solution that will be realized in practice 
will depend on the initial and boundary conditions of the full time-dependent system.

Here we consider a zonal torque of the form
\begin{eqnarray}\label{aforce}
{\cal F} = & {\cal F}_0 (r^2 + a r + b) r \sin^2\theta \cos^2\theta & \mbox{\hspace{.5in}(Region 1: $r_s \leq r \leq R$)} \\
         = & 0                              & \mbox{\hspace{.5in}(Region 2: $r < r_s$, $z \geq z_e$)} \\
         = & {\cal F}_e(r,\theta) & \mbox{\hspace{.5in}(Region 3: $z < z_e$)}
\label{aforce3}
\end{eqnarray}
where ${\cal F}_0$ is the amplitude of the force and $r_s$ is the
bottom of the NSSL (Region 1).  The coefficients $a$ and $b$ will 
be chosen to ensure that the meridional flow is continuous across $r_s$.  
Also, we'll choose $a$, $b$, and ${\cal F}_0$ to give us a negative 
${\cal F}$ in Region 1.  The latitudinal dependence in (\ref{aforce}) is 
chosen to produce a $v_\theta$ that peaks at mid-latitudes, going as
$\sin\theta \cos\theta$ in the limit $\Delta \Omega \rightarrow 0$
(see eq.\ (\ref{eq:vtheta}) below).  The boundary layer at the equator 
(Region 3, equation (\ref{aforce3})] is included to close the circulation 
cell in the northern hemisphere and to ensure that the net torque 
$\int_V {\cal F} dV = 0$.  Since is it a passive response, we will compute 
it only after we have computed the mean flow in Regions 1 and 2.  The
free parameters that specify ${\cal F}$ are thus the amplitude in the
NSSL, ${\cal F}_0$, and the locations of the boundary layers, $r_s$
and $z_e$.

From equation (\ref{aomega}) we have
\begin{equation}\label{dLdlam}
\frac{d{\cal L}}{d\lambda} = 2 \Omega_0 \lambda \left(1 + \gamma \lambda\right)
\end{equation}
where $\gamma = 3\Delta \Omega/(2\Omega_0 R)$  ~~~.

Equation (\ref{eq:gp}) yields the $\uvl$ component of the flow in Region 1:
\begin{equation}\label{avlam}
\left<\rho v_\lambda\right> = \beta ~ \frac{r^2 + ar + b}{1 + \gamma \lambda} ~ \frac{\lambda z^2}{r^3}\mbox{\hspace{.5in}(Region 1)}
\end{equation}
where $\beta = {\cal F}_0/(2 \Omega_0)$.  In obtaining (\ref{avlam}) we have used the relations 
$\lambda = r \sin\theta$ and $z = r \cos\theta$.  

Equation (\ref{eq:psisoln}) then yields
\begin{eqnarray}\label{psiappx}
\Psi(\lambda,z^\prime) = & \frac{\beta \lambda}{1 + \gamma \lambda} 
\left[I(\lambda,z) - I(\lambda,z_b)\right]
& \mbox{\hspace{.5in}($z_s \leq z^\prime \leq z_b$)} \\
\Psi(\lambda,z^\prime) = & \Psi(\lambda,z_s) & \mbox{\hspace{.5in}($z_e \leq z^\prime < z_s$)} 
\nonumber \\
\Psi(\lambda,z^\prime) = & \Psi_3(\lambda,z) & \mbox{\hspace{.5in}($0 \leq z^\prime < z_e$)} 
\nonumber
\end{eqnarray}
where $z_b = \sqrt{R^2 - \lambda^2}$ as in \S\ref{sec:gp}, $z_s = \sqrt{r_s^2 - \lambda^2}$,
and we will specify $\Psi_3(\lambda,z)$ below.  The solution in Region 1 involves the
following integral
\begin{equation}\label{Iz}
I(\lambda,z) = \int \left(r^2 + a r + b\right) ~ \frac{z^2}{r^3} ~ dz =
\frac{z}{2 r} \left(r^2 + 2 a r - 2 b\right) + 
\left(b - \frac{\lambda^2}{2}\right) \ln \vert z + r \vert
- a \lambda \tan^{-1}\left(\frac{z}{\lambda}\right)  ~~~. 
\end{equation}
We have dropped the integration constant because we'll use this as a definite 
integral in what follows.  

In order to compute the coefficients $a$ and $b$
we will need the axial flow component, which follows from equation (\ref{eq:psidef})
\begin{equation}\label{vzsoln1}
\left<\rho v_z\right> = - \beta ~ \frac{2 + \gamma \lambda}{\left(1 + \gamma \lambda\right)^2} 
\left[I(\lambda,z) - I(\lambda,z_b)\right]
- \frac{\beta \lambda}{1 + \gamma \lambda} 
\left[\frac{\pd}{\pd \lambda} I(\lambda,z) - 
\frac{\pd }{\pd \lambda} I(\lambda,z_b)\right]  
\mbox{\hspace{.5in}(Region 1) ,}
\end{equation}
with $\left<\rho v_z\right> = \left<\rho v_z\right>\vert_{z=z_s}$ in Region 2. 

Note that the term in equation (\ref{vzsoln1}) involving 
$I(\lambda,z) - I(\lambda,z_b)$ vanishes at $z = z_b$ but as we'll
see below, the term involving their derivatives does not.  This is
consistent with the impenetrable boundary condition $v_r = 0$ at 
$r = R$ since the nonzero $v_\theta$ has both $v_\lambda$ and $v_z$
components.

The derivatives in equation (\ref{vzsoln1}) are given by
\begin{equation}\label{dIdlam}
\frac{\pd }{\pd \lambda} I(\lambda,z) = \left(r^2 + 2 a r + 2b\right) \frac{\lambda z}{2 r^3}
- \lambda \ln \vert z + r \vert + \frac{2 b - \lambda^2}{z + r} ~ \frac{\lambda}{2 r}
- a \tan^{-1}\left(\frac{z}{\lambda}\right) 
\end{equation}
and
\begin{equation}\label{dIbdlam}
\frac{d }{d \lambda} I(\lambda,z_b) = \left(2 b - R^2\right) \frac{\lambda}{2 R z_b} 
- \lambda \ln \vert z_b + R\vert - \frac{2 b - \lambda^2}{z_b + R} ~ \frac{\lambda}{2 z_b}
- a \tan^{-1}\left(\frac{z_b}{\lambda}\right)  ~~~,
\end{equation}
where we have used $d z_b/d\lambda = - \lambda / z_b$.

Now we proceed to compute the coefficients $a$ and $b$.  Since $v_\lambda = 0$
for $r < r_s$, then continuity of $v_\lambda$ requires that it should vanish
at $r = r_s$ as well, along with the forcing ${\cal F}$.  Using (\ref{aforce}), this
can be achieved if $b = - r_s \left(r_s + a\right)$.

As a consequence of the way we set up the problem, we also have
\begin{equation}\label{divbc}
\frac{\pd}{\pd \lambda} \left(\lambda \rho v_\lambda\right) =
\frac{\pd }{\pd z} \left(\rho v_z\right) = 0  
\mbox{\hspace{.5in} (Region 2)}  ~~~.
\end{equation}
So, in order for the circulation to be continuous, we require these
expressions to hold at $r = r_s$ as well.  It is straightforward to
show from equation (\ref{avlam}) that equation (\ref{divbc}) is 
satisfied at $r = r_s$ if $r_s^2 + 2 a r_s + 3 b = 0$.
Combining this with the previous expression then yields
$a = - 2 r_s$ and $b = r_s^2$.

Note that with these values of $a$ and $b$, we have
$r^2 + a r + b = \left(r - r_s\right)^2$.
So, if we choose a negative value for ${\cal F}_0$
(which implies $\beta < 0$), then equation (\ref{aforce})
ensures that ${\cal F} < 0$ in the NSSL, as desired.

\begin{figure}
\centerline{\epsfig{file=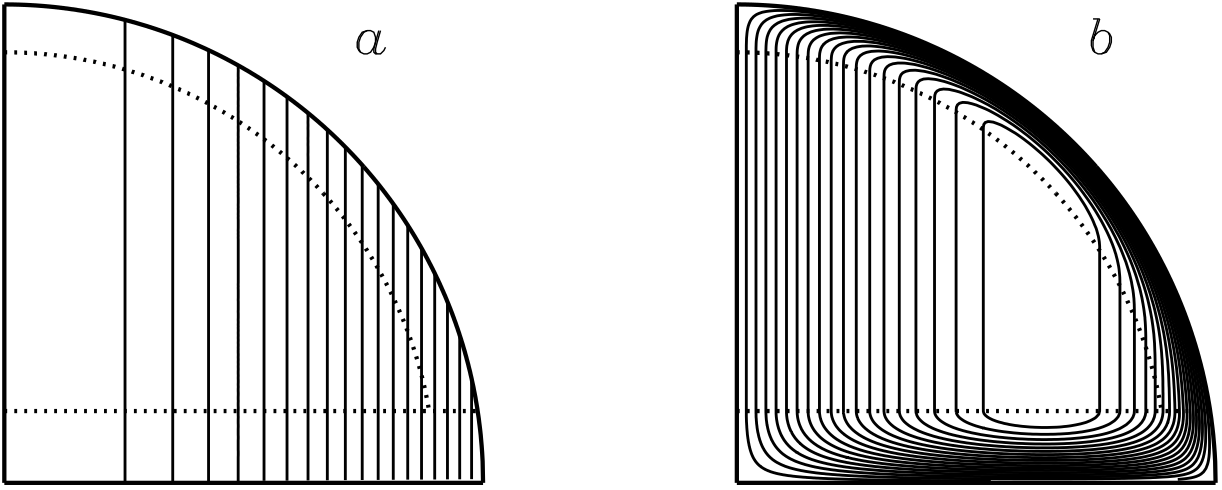,width=\linewidth}}
\caption{The analytic solution for ($a$) ${\cal L}$ and ($b$) $\Psi$ 
according to equations (\ref{aomega}) and 
(\ref{psiappx}).  Here the NSSL is a bit wide, $r_s = 0.90 R$
(indicated by the dotted line), in order to highlight the structure of 
the boundary layer.  Here we've taken $\gamma = 0.45 R^{-1}$ and
$z_e = 0.15 R$.  Contours in ($b$) represent streamlines of the mass flux, 
with poleward flow at the surface.  Dotted lines delineate Regions 1, 2, 
and 3.\label{gyro_mc}}
\end{figure}

This completes the solution for $\Psi(\lambda,z)$ in Regions 1 and 2.
For Region 3 we can then set
\begin{equation}\label{psi3}
\Psi_3(\lambda,z) = \frac{z}{z_e^2} \left(2 z_e - z\right) ~ \Psi(\lambda,z_e)  
+ \frac{z}{z_e} \left(z - z_e\right) \Psi^\prime(\lambda,z_e) ~~~,
\end{equation}
where $\Psi^\prime = \pd \Psi/\pd z = \left<\rho v_\lambda\right>$. The
value at $z = z_e$ is zero for $r_e < r_s$ and is given by 
equation (\ref{avlam}) for $r_e > r_s$.  Here $r_e = (z_e^2 + \lambda^2)^{1/2}$.
It is straightforward to show that this yields $\Psi = 0$ at $z = 0$, implying no
flow across the equatorial plane, and furthermore, that $\Psi$, $\pd \Psi/\pd \lambda$,
and $\pd \Psi / \pd z$ are continuous across $z = z_e$.  The analytic solution is
plotted in Figure \ref{gyro_mc}.

The corresponding cylindrically outward flow in Region 3 is given by equation (\ref{eq:psidef})
\begin{equation}\label{rv3}
\left<\rho v_\lambda\right> = \frac{2}{z_e^2} \left(z_e-z\right) \Psi(\lambda,z_e) 
+ \frac{2 z - z_e}{z_e} \Psi^\prime(\lambda,z_e) 
\mbox{\hspace{.5in} (Region 3)}  ~~~,
\end{equation}
while equation (\ref{eq:gp}) gives the required torque
\begin{equation}\label{torque3}
{\cal F}_3(\lambda,z) = \frac{d {\cal L}}{d \lambda} \left<\rho v_\lambda\right>
\end{equation}
with $d {\cal L}/d\lambda$ given by (\ref{dLdlam}) and $\left<\rho v_\lambda\right>$ 
given by (\ref{rv3}).  Note that the volume-integrated torque is zero by construction,
as a consequence of the impenetrable boundaries and equation (\ref{eq:gp})
\begin{equation}
\int_V {\cal F} dV = \int_V \left<\rho \vv_m\right> \bdot \del {\cal L} dV
= \int_S \dv \left(\left<\rho \vv_m\right> {\cal L}\right) \bdot d{\bf S} = 0 ~~~.
\end{equation}

The axial flow in Region 3 follows from equations (\ref{eq:psidef}),
(\ref{vzsoln1}), and (\ref{psi3}) 
\begin{eqnarray}
\left<\rho v_z\right> = & \frac{z}{z_e^2} \left(2 z_e - z\right) 
\left.\left<\rho v_z\right>\right\vert_{z=z_s} 
& \mbox{\hspace{.2in} (Region 3, $r \leq r_s$)}  \nonumber \\ 
\left<\rho v_z\right> = & \frac{z}{z_e^2} \left(2 z_e - z\right)
\left.\left<\rho v_z\right>\right\vert_{z=z_e} 
- \beta \frac{z^3 \lambda}{r^3 z_e} \frac{(r-r_s)^2}{1 + \gamma \lambda}
\left(z - z_e\right) \left(\frac{2}{r-r_s} + \frac{2}{\lambda} 
- \frac{\gamma}{1 + \gamma \lambda} - \frac{3 \lambda}{r^2}\right)
& \mbox{\hspace{.2in} (Region 3, $r > r_s$)} \nonumber ~~~.
\end{eqnarray}

It is instructive to verify that the flow at the outer boundary is indeed
poleward.  Substituting $z = z_b$ and $r = R$ into equations (\ref{avlam}) 
and (\ref{vzsoln1}) and expressing the result in spherical coordinates
yields
\begin{equation}\label{rvlamb}
\left<\rho v_\lambda\right> = \frac{\beta D^2}{1 + \gamma R \sin\theta} ~ \sin\theta \cos^2\theta
\mbox{\hspace{.1in}, and \hspace{.1in}}
\left<\rho v_z\right> = - \frac{\beta D^2}{1 + \gamma R \sin\theta} ~ \sin^2\theta \cos\theta
\mbox{\hspace{.2in}($r = R$)}
\end{equation}
where $D = R - r_s$ is the thickness of the NSSL.  Furthermore,
\begin{equation}\label{eq:vtheta}
\left<\rho v_\theta\right> = -\sin\theta \left<\rho v_z\right> + \cos\theta \left<\rho v_\lambda\right> = 
\frac{{\cal F}_0 D^2}{2 \Omega_0} ~ 
\frac{\sin\theta \cos\theta}{1 + (\Delta \Omega/\Omega_0) \sin\theta} 
\mbox{\hspace{.5in}($r = R$) ~~~,}
\end{equation}
and $\left<\rho v_r\right> = \cos\theta \left<\rho v_z\right> + \sin\theta \left<\rho v_\lambda\right> = 0$ at $r=R$.
For ${\cal F}_0 < 0$, equation (\ref{eq:vtheta}) implies a poleward flow.

This is clearly a highly idealized depiction of a star.  In the solar
convective envelope, strong turbulent stresses ${\cal F}$ (and possibly 
${\cal G}$) and baroclinic forcing ${\cal B}$ maintain a substantial
non-cylindrical differential rotation profile (Fig.\ \ref{fig:DR}) and
this will in turn determine how the meridional circulation streamlines
close in Regions 2 and 3. In particular, the initial circulations 
established by convection on a dynamical time scale are unlikely to 
penetrate much below the convection zone \citep{gilma04}, although
very weak gyroscopically-pumped circulations driven by convection
will burrow downward on a radiative diffusion time scale, eventually
reaching the deep radiative interior over the course of billions
of years \citep{spieg92,gough98,garau08}.  Here turbulent stresses
are weak ${\cal F} \approx 0$ and the rotation is nearly uniform
so the circulation contours would follow cylindrical surfaces as
in Region 2 of the present example.  In any case, the solution 
discussed in this section is only intended to give a rough feel 
for how gyroscopic pumping in the NSSL might impact the dynamics 
in the upper convection zone.

Still, equations (\ref{aomega}), (\ref{psiappx}), (\ref{Iz}),
and (\ref{psi3}). provide a steady, analytic solution
to the continuity, momentum, and energy equations in the barotropic
(or isentropic) limit ($P = P(\rho)$) under the influence of specified 
turbulent stresses analogous to the NSSL (${\cal F} < 0$ for $r > 0.95$).
They also highlight the need for turbulent or baroclinic stresses in
the meridional plane, ${\cal B}$ and/or ${\cal G}$, in order to 
maintain an axial rotational shear $\pd \Omega/\pd z \neq 0$.

\section{Appendix C: Uncurling the Zonal Vorticity Equation}

As described in \S\ref{sec:anen}, we wish to derive the acceleration ${\bf A}_m$
corresponding to the meridional stress ${\cal G}$.  To do this, we must solve
equation (\ref{eq:zetaeq}) for $\zeta$.  The expansion for $\Omega$ in 
equation (\ref{eq:omega}) implies that we can write $\zeta$ as
\begin{equation}
\zeta(r,\theta) = \sin\theta \sum_n Z_n(r) \cos^n\theta
\mbox{\hspace{.2in} ($n =$ 1, 3, 5, 7, 9).}
\end{equation}
Each of the $Z_n(r)$ satisfy a recursive equation of the form
\begin{equation}\label{eq:Zrec}
Z_n^{\prime\prime} + \frac{2}{r} Z_n^\prime - \frac{1}{r^2}
\left(n^2+ 3 n + 2\right) Z_n = - \Gamma_n - \frac{1}{r^2} \left(n+1\right) 
\left(n+2\right) Z_{n+2}
\end{equation}
where primes denote (ordinary) derivatives with respect to $r$ and
$Z_n = 0$ for $n$ even and for $n > 9$.  The $\Gamma_n$
are given by
\begin{equation}
\Gamma_1 = - 2 r \Omega_e d_1 \mbox{\hspace{.2in}} 
\Gamma_3 = - 2 r \left(\Omega_e d_3 + \Omega_2 d_1\right) \mbox{\hspace{.2in}} 
\Gamma_5 = - 2 r \left(\Omega_e d_5 + \Omega_2 d_3 + \Omega_4 d_1\right)
\end{equation}
\begin{equation}
\Gamma_7 = - 2 r \left(\Omega_2 d_5 + \Omega_4 d_3\right) \mbox{\hspace{.5in}} 
\Gamma_9 = - 2 r \Omega_4 d_5 \mbox{\hfill}
\end{equation}
where
\begin{equation}\label{eq:dees}
d_1 = \Omega_e^\prime + \frac{2}{r} \Omega_2 \mbox{\hspace{.3in}} 
d_3 = \Omega_2^\prime - \frac{2}{r} \Omega_2 + \frac{4}{r} \Omega_4 \mbox{\hspace{.3in}} 
d_5 = \Omega_4^\prime + \frac{2}{r} \Omega_2 - \frac{4}{r} \Omega_4 ~~~.
\end{equation}
The radial derivatives in equations (\ref{eq:Zrec}) and (\ref{eq:dees}) are discretized
using a second-order finite difference scheme and the resulting matrix equations
are solved subject to the boundary conditions $Z_n = 0$ at $r = R$ and
$r Z^\prime_n + Z_n = 0$ at $r = r_s$.  These boundary conditions ensure that 
the vertical acceleration vanishes at the surface ($A_r = 0$ at $r = R$) 
and the horizontal acceleration vanishes at the base of the NSSL
($A_\theta = 0$ at $r = r_s$).

\section{Appendix D: Diffusive Angular Velocity Profiles}

In this Appendix we seek angular velocity profiles for which the net axial torque
vanishes ${\cal F} = 0$, subject to the boundary conditions
\begin{equation}\label{eq:omega_m}
\Omega(r_s,\theta) = \hat{\Omega}_0 + \hat{\Omega}_2 \cos^2\theta + 
\hat{\Omega}_4 \cos^4\theta =
\omega_0 + \omega_2 P_2(\cos\theta) + \omega_4 P_4(\cos\theta)
\end{equation}
and
\begin{equation}
\left. \frac{\pd \Omega}{\pd r}\right\vert_{r=r_s} = 0 
\end{equation}
at a specified matching layer, $r=r_s$.  Here $\hat{\Omega}_0$, $\hat{\Omega}_2$ 
and $\hat{\Omega}_4$ are fitting coefficients corresponding to the solar rotation 
profile at the matching layer ($r = r_s$), as inferred from global
helioseismology (Fig.\ \ref{fig:DR}).  The expression on the far right 
of equation (\ref{eq:omega_m}) provides an alternate representation of 
$\Omega(r_s,\theta)$ in terms of Legendre Polynomials $P_n(x)$.  The
Legendre coefficients are given by 
\begin{equation}
\omega_0 = \hat{\Omega}_0 + \frac{\hat{\Omega}_2}{3} + \frac{\hat{\Omega}_4}{5}
\mbox{\hspace{.5in}}
\omega_2 = \frac{2}{3} \hat{\Omega}_2 + \frac{4}{7} \hat{\Omega}_4
\mbox{\hspace{.5in}}
\omega_4 = \frac{8}{35} \hat{\Omega}_4 ~~~.
\end{equation}

We choose $r_s$ to correspond to the location where the radial gradient of 
the spherically-averaged angular velocity profile 
passes through zero.  This yields values of 
$r_s = 0.946 R$, $\hat{\Omega}_0/(2\pi) = $ 468 nHz, $\hat{\Omega}_2/(2\pi) = $ - 62.5 nHz, 
and $\hat{\Omega}_4/(2\pi) = $ - 77.9 nHz.

We consider two hypothetical forms for ${\cal F}$, representing 
turbulent diffusion and the mixing of angular momentum
as discussed in \S\ref{sec:transport}.  In the first case the 
angular momentum flux is given by equation (\ref{eq:Fvd}).
The assumption of no net torque and a constant density-weighted
diffusion coefficient $\rho \nu_t$ then yields 
equation (\ref{eq:vdeq}): $\del (\lambda^2 \del\Omega_v) = 0$.  
The solution can be written as a Legendre series
\begin{equation}\label{eq:omega_t}
\Omega_v(r,\theta) = W_0(r) + W_2(r) P_2(\cos\theta) + W_4(r) P_4(\cos\theta)
\end{equation}
with coefficients
\begin{equation}
W_4(r) = a_1 \left(r^4 + a_2 r^{-7}\right)
\end{equation}
\begin{equation}
W_2(r) = b_1 r^4 + b_2 r^{-7} + b_3 r^2 + b_4 r^{-5} 
\end{equation}
\begin{equation}
W_0(r) = c_1 r^4 + c_2 r^{-7} + c_3 r^2 + c_4 r^{-5} 
+ c_5 r^{-3} + c_6
\end{equation}
where 
\begin{equation}
a_2 = \frac{4}{7} r_m^{11} 
\mbox{\hspace{.5in}}
a_1 = \frac{\omega_4}{r_m^4+a_2 r_m^{-7}}
\mbox{\hspace{.5in}}
b_1 = \frac{5}{9} a_1 
\mbox{\hspace{.5in}}
b_2 = \frac{5}{9} a_1 a_2
\end{equation}
\begin{equation}
b_3 = \frac{5}{7} \omega_2 r_m^{-2} - \frac{9}{7} b_1 r_m^2 + \frac{2}{7} b_2 r_m^{-9}
\mbox{\hspace{1.5in}}
b_4 = \omega_2 r_m^5 - b_1r_m^9 - b_2 r_m^{-2} - b_3 r_m^7
\end{equation}
\begin{equation}
c_1 = \frac{a_1+b_1}{14}  \mbox{\hspace{.5in}}
c_2 = \frac{a_1 a_2 + b_2}{14} \mbox{\hspace{.5in}}
c_3 = \frac{b_3}{5}   \mbox{\hspace{.5in}}
c_4 = \frac{b_4}{5}
\end{equation}
\begin{equation}
c_5 = \frac{1}{3} \left(4 c_1 r_m^7 - 7 c_2 r_m^{-4} + 2 c_3 r_m^5 - 5 c_4 r_m^{-2}\right)
\hspace{.4in} 
c_6 = \omega_0 - c_1 r_m^4 - c_2 r_m^{-7} - c_3 r_m^2 - c_4 r_m^{-5} - c_5 r_m^{-3} ~~~.
\end{equation}

For the second case we consider, that of angular momentum mixing, the angular 
momentum flux is given by equation (\ref{eq:Fmx}) and the rotation profile
for ${\cal F} = 0$ and $\rho \nu_a$ = constant is given by
equation (\ref{eq:mxeq}), $\nabla^2{\cal L}_a = 0$.
The solution is readily obtained through a Legendre series
\begin{equation}\label{eq:omega_a}
{\cal L}_a(r,\theta) = \sum_n 
\frac{r^n + \beta_n r^{-(n+1)}}{r_m^n + \beta_n r_m^{-(n+1)}} 
~ L_n P_n(\cos\theta)
\mbox{\hspace{.3in} ($n = $ 0, 2, 4, 6).}
\end{equation}
Here $L_n$ are the coefficients corresponding to the matching layer
\begin{equation}
L_6 = - \frac{16}{231} ~ \hat{\Omega}_4 r_m^2
\hspace{.5in}
L_4 = \frac{8}{35} \left(\hat{\Omega}_4-\hat{\Omega}_2\right) r_m^2
+ \frac{9}{2} L_6
\end{equation}
\begin{equation}
L_2 = \frac{2}{3}\left(\hat{\Omega}_2-\hat{\Omega}_0\right) r_m^2 
+ \frac{5}{2} L_4 - \frac{35}{8} L_6
\hspace{.5in}
L_0 = \hat{\Omega}_0 r_m^2 + \frac{L_2}{2} - \frac{3}{8} L_4
+ \frac{5}{16} L_6
\end{equation}
and $\beta_n = \left[(n - 2)/(n + 3)\right] ~ r_m^{2 n + 1}$.
The corresponding rotation profile is then $\Omega_a = \lambda^{-2} {\cal L}_a$. 

The two solutions derived here, $\Omega_v(r,\theta)$ and $\Omega_a(r,\theta)$
are shown in Figure \ref{fig:dp}.

\end{document}